\documentclass[nofootinbib,superscriptaddress,aps]{revtex4}
\usepackage{amssymb,amsmath,graphicx,bm,color}
\usepackage[caption=false]{subfig}

\renewcommand{\d}{\mathrm{d}}

\newcommand{\bfp}{\mbox{\boldmath $p$}}
\newcommand{\pup}{p^\uparrow}

\newcommand{\bfk}{\mbox{\boldmath $k$}}

\newcommand{\ks}{k \!\!\! /}
\newcommand{\ps}{p \!\!\! /}

\newcommand{\pgs}{p_g  \!\!\!\!\!\! / ~}
\newcommand{\epss}{\varepsilon\!\!\!/}

\newcommand{\pqs}{{p_{\cal Q}} \!\!\!\!\!\!\! /~\,\,}

\def\slash#1{\setbox0=\hbox{$#1$}               
        \dimen0=\wd0                            
        \setbox1=\hbox{/} \dimen1=\wd1          
        \ifdim\dimen0>\dimen1                   
        \rlap{\hbox to \dimen0{\hfil/\hfil}}    
        #1                                      
        \else
        \rlap{\hbox to \dimen1{\hfil$#1$\hfil}} 
        /                                       
        \fi}                                    %

\begin{document}

\title{Probing the gluon Sivers function in $p^\uparrow p\to J/\psi\,X$ and $p^\uparrow p \to D\,X$}

\author{Umberto D'Alesio}
\email{umberto.dalesio@ca.infn.it}
\affiliation{Dipartimento di Fisica, Universit\`a di Cagliari, Cittadella Universitaria, I-09042 Monserrato (CA), Italy}
\affiliation{INFN, Sezione di Cagliari, Cittadella Universitaria, I-09042 Monserrato (CA), Italy}

\author{Francesco Murgia}
\email{francesco.murgia@ca.infn.it}
\affiliation{INFN, Sezione di Cagliari, Cittadella Universitaria, I-09042 Monserrato (CA), Italy}

\author{Cristian Pisano}
\email{cristian.pisano@unipv.it}
\affiliation{Dipartimento di Fisica, Universit\`a di Pavia, via Bassi 6, I-27100 Pavia, Italy}
\affiliation{INFN Sezione di Pavia, via Bassi 6, I-27100 Pavia, Italy}

\author{Pieter Taels}
\email{pieter.taels@uantwerpen.be}
\affiliation{Department of Physics, University of Antwerp, Groenenborgerlaan 171, B-2020 Antwerp, Belgium}

\begin{abstract}
We present a study of transverse single-spin asymmetries (SSAs) in $p^\uparrow p\to J/\psi\,X$  and $p^\uparrow p\to D X$ within the framework of the generalized parton model (GPM), which includes both spin and transverse momentum effects, and show how they can provide useful information on the still almost unknown gluon Sivers function. Moreover, by adopting a modified version of this model, named color gauge invariant (CGI) GPM, we analyze the impact of the initial- and final-state interactions on our predictions. As a consequence, we find that these two processes are sensitive to different gluon Sivers functions, which can be expressed as linear combinations of two distinct, universal gluon distributions. We therefore define proper observables which could allow for a separate extraction of these two independent Sivers functions. At the same time, we show how it would be possible to discriminate between the GPM and the CGI-GPM approaches by comparing the corresponding estimates of SSAs with present and future experimental results at RHIC.
\end{abstract}

\date{\today}

\maketitle

\section{Introduction}

Transverse single-spin asymmetries (SSAs) in high-energy lepton-hadron and hadron-hadron reactions are an invaluable tool to probe hadrons at a deeper level of accuracy, as well as to get information on the intimate role of strong interactions. Indeed, they can provide information on the three-dimensional structure of the nucleons and at the same time  shed light, or at least give some hints, on the still-unknown mechanism of confinement.

SSAs are defined as the ratio of the difference and sum of cross sections in which the spin of one of the hadrons is reversed. They have stimulated the research program of many experiments, like HERMES at HERA (DESY), and, more recently, COMPASS at CERN, CLAS and CLAS12 at Jefferson Lab and STAR, PHENIX and BRAHMS at RHIC (Brookhaven National Laboratory): see for instance Refs.~\cite{Aschenauer:2015ndk,Barone:2010zz,D'Alesio:2007jt} for recent reviews.

{}From the theoretical point of view, the interpretation of SSAs within the framework of the common leading-twist, collinear factorization theorems in QCD is very challenging. At present, essentially two approaches have been proposed, and are under current investigation, to explain such effects.

One, based on the factorization of hard scattering cross sections and transverse-momentum-dependent (TMD) parton distribution and fragmentation functions (PDFs and FFs), is proven to be valid for processes characterized by two energy scales~\cite{Collins:1984kg,Ji:2004xq,Collins:2011zzd,GarciaEchevarria:2011rb}: a hard one, like the virtuality of the exchanged boson in semi-inclusive deep inelastic scattering (SIDIS), Drell-Yan (DY) processes or $e^+e^-$ annihilation, and a soft one, of the order of $\Lambda_{\rm QCD}$, like the transverse momentum of the final hadron in SIDIS, or of the lepton pair in DY, or the transverse momentum imbalance in hadron-pair production in $e^+e^-$ collisions. From the phenomenological point of view, a prominent example of a TMD-FF is provided by the Collins function~\cite{Collins:1992kk}, describing the fragmentation of a transversely polarized quark into a noncollinear unpolarized hadron. Among the TMD-PDFs, the Sivers function~\cite{Sivers:1989cc} represents the azimuthal distribution of unpolarized quarks and gluons in a nucleon polarized transversely to its direction of motion. In contrast to FFs, supposed to be universal, TMD-PDFs are intrinsically process dependent because of the effects of initial- and final-state interactions (ISIs and FSIs), encoded in the Wilson lines (or gauge links) entering  their color gauge-invariant definition. A typical example is provided by the predicted sign change of the quark Sivers function in SIDIS and in DY processes, due to the presence of FSIs and ISIs,  respectively~\cite{Collins:2002kn,Brodsky:2002rv}. In those processes where both ISIs and FSIs contribute, the color structure of the Sivers function is more complicated and TMD factorization could even be broken~\cite{Rogers:2010dm}.

On the other hand, a second approach, based on collinear factorization at next-to-leading twist (twist-three), is suitable for the description of processes characterized by only one hard energy scale, {\it i.e.}\ the transverse momentum of a particle inclusively produced in hadronic collisions~\cite{Efremov:1984ip,Qiu:1991pp,Qiu:1991wg,Kouvaris:2006zy,Kanazawa:2014dca}. In this framework, SSAs are given by convolutions of hard scattering amplitudes with universal quark-gluon-quark and three-gluon correlation functions.

In a series of papers~\cite{Anselmino:1994tv, D'Alesio:2004up, Anselmino:2005sh, Anselmino:2011ch, Anselmino:2012rq, Anselmino:2013rya,D'Alesio:2010am}, even if not supported by a formal proof, the validity of the TMD formalism has been assumed for single-scale processes as well, like $p^\uparrow p\to \pi\,X$. Moreover, in such a scheme, TMD-PDFs are conditionally taken to be universal. This phenomenological approach is nowadays known as generalized parton model (GPM) and is able to successfully describe many features of several available data. More recently, the process dependence of the {\em quark} Sivers function has been studied within the GPM for proton-proton collisions, by taking into account the effects of ISIs and FSIs under a one-gluon exchange approximation, leading to the so-called color gauge invariant formulation of the GPM, referred to as CGI-GPM~\cite{Gamberg:2010tj,D'Alesio:2011mc,D'Alesio:2013jka}. In this approach, the process dependence of the quark Sivers function can be shifted to the partonic cross sections. Therefore the Sivers function can still be considered universal, but when calculating physical observables it has to be convoluted with modified partonic cross sections, which turn out to have the same form, in terms of Mandelstam variables, of the hard functions of the twist-three collinear approach~\cite{Gamberg:2010tj}. In particular, this model is able to reproduce the expected opposite relative sign of the Sivers asymmetries for SIDIS and DY, due to the effects of FSIs and ISIs, respectively~\cite{Collins:2002kn,Brodsky:2002rv}.

Along the same lines, here we address two aspects that, as we are going to show, are somehow related to each other. From one side we will focus on how to get information on an important TMD-PDF so far poorly explored, namely the gluon Sivers function (GSF)~\cite{Boer:2015vso}. To this end we consider SSAs in hadronic processes where its contribution is expected to be dominating. On the other hand, in the spirit of pursuing and deepening the study of the process dependence of the Sivers function, we compute color-gauge initial- and final-state interactions for these same processes (that are characterized by the presence of only one large energy scale).

We will then show, within the frameworks of both the GPM and CGI-GPM, how the analysis of existing and future data for SSAs in the single-polarized processes $p^\uparrow p\to J/\psi \,X$ and $p^\uparrow p\to D\,X $ could constrain the
gluon Sivers function. To this end, for the first time, we extend the methods developed in Ref.~\cite{Gamberg:2010tj} to the gluon sector. As for the quark case, the process dependence of the gluon Sivers function can be absorbed into the partonic hard functions. However, one has to introduce two universal, completely independent, Sivers distributions because, for three colored gluons, there are two different ways of forming a color-singlet state. The totally antisymmetric color combination, even under charge conjugation, is commonly referred to as an $f$-type state, while the symmetric combination, odd under $C$-parity, is referred to as a $d$-type state. Hence, in analogy to Ref.~\cite{Bomhof:2006ra}, we introduce an $f$-type and a $d$-type gluon Sivers function, which are named $A_1$ and $A_2$ in the notation of Ref.~\cite{Buffing:2013kca} and are related to the two distinct trigluon Qiu-Sterman functions in the collinear, twist-three formalism~\cite{Ji:1992eu,Kang:2008qh,Kang:2008ih}, as will be discussed also in the following. A similar analysis within only the GPM approach has been presented in Refs.~\cite{Godbole:2016tvq,Godbole:2017syo}

We note that a previous extraction of the gluon Sivers function from $p^\uparrow p \to \pi^0X $ at central rapidities~\cite{D'Alesio:2015uta} assumed the universality of this distribution, as it is in the GPM approach. A reanalysis of those data within the CGI-GPM is on the way~\cite{dalesio2017}. As compared to the pion case, inclusive $J/\psi$ and $D$ production have the advantage of probing gluon TMDs directly, since the contributions from quark-initiated processes turn out to be negligible. In the present study, therefore, in order to show an independent, direct and unbiased way to gather information on this TMD,
we will not make use of the results obtained in Ref.~\cite{D'Alesio:2015uta}.

For completeness we mention that the gluon Sivers function could be probed as well in the process $\ell \,p^\uparrow \to \ell^\prime J/\psi \,X$~\cite{Godbole:2012bx,Godbole:2013bca,Mukherjee:2016qxa}, for which only preliminary data are currently available from the COMPASS Collaboration at CERN~\cite{Matousek:2016xbl}. Moreover, the COMPASS Collaboration has started analyzing the gluon Sivers effect in the production of high-$p_T$ hadron pairs in muon scattering off polarized proton and deuteron targets~\cite{Szabelski:2015sza,Adolph:2017pgv}. At a future Electron-Ion Collider, the study of SSAs for $ep^\uparrow \to e^\prime Q\,\overline{Q}\,X$ would provide information on the $f$-type gluon Sivers function~\cite{Boer:2016fqd}. This distribution could also be accessed at RHIC by looking at diphoton production~\cite{Qiu:2011ai}, while $p^\uparrow p \to \gamma\, {\rm jet}\, X$ would be sensitive to the $d$-type gluon distribution~\cite{Boer:2015vso,Boer:2016bfj}. Other promising measurements of the ($f$-type) gluon Sivers function could be performed at a future fixed target experiment at the LHC, named AFTER@LHC, and include processes like $p^\uparrow p \to \eta_{c,b}\,X$ and $p^\uparrow p \to J/\psi (\Upsilon)\, \gamma\,X$ for which it has been shown that the color-singlet production mechanism dominates~\cite{Boer:2012bt,Dunnen:2014eta}.

The remainder of the paper is organized as follows: in Sec.~\ref{jpsi} we consider $J/\psi$ production in (un)polarized proton-proton collisions; in particular, in Sec.~\ref{csm} we present our results in the color-singlet model within a TMD approach, in Sec.~\ref{SSAjpsi} the calculation of the contribution of initial- and final-state interactions to the SSA and then in Sec.~\ref{Jpsi-res} our theoretical estimates compared with available data. We then move in Sec.~\ref{SSA-D} to the SSAs for $D$ meson production in proton-proton collisions. More precisely, in Sec.~\ref{Dmes-form} we discuss and present some details of the theoretical calculations and in Sec.~\ref{Dmes-res} we collect our main results. Conclusions and open issues are then gathered in Sec.~\ref{concl}. We collect some further theoretical details in Appendixes~\ref{app:unp-J} and~\ref{app:CGI-D}.

\section{Cross sections and single spin asymmetries for $p p \to J/\psi\,\,X$}
\label{jpsi}

\subsection{Production mechanism: The color-singlet model}
\label{csm}

We consider first the inclusive production of a quarkonium state $\cal Q$ in unpolarized proton-proton scattering,
\begin{equation}
p(p_A)\,{+}\,p(p_B)\,\to\, {\cal Q} (p_{\cal Q}) \, {+}  \,X \, ,
\label{eq:proc-Jpsi}
\end{equation}
where the four-momenta of the particles are given within parentheses. We assume that the colorless heavy quark-antiquark pair forming the quarkonium  is in a bound state described by a nonrelativistic wave function  with spin $S=1$, orbital angular momentum $L=0$ and total angular momentum $J=1$. In the following we adopt  the spectroscopic notation ${\cal Q} \equiv Q \overline Q[^{2S+1}L_J^{(1,8)} ]$, where the color assignments for the quark pair are generally specified by the singlet or octet superscripts, $(1)$ or $(8)$. Therefore, in our case,  $ {\cal Q} = Q \overline{Q}[ ^3S_1^{(1)} ] $ with $Q=c,b$. The squared invariant mass of the resonance is denoted by $M^2=p_{\cal Q}^2$, with $M$ being twice the heavy quark mass up to small relativistic corrections.

Within the framework of the CSM (see e.g.~Ref.~\cite{Baier:1983va}), the heavy quark and antiquark pair is produced in the hard partonic scattering with the same quantum numbers as the meson into which it nonperturbatively evolves. Therefore, $J/\psi$ production is dominated, at leading order (LO) $\alpha_s^3$ in perturbative QCD, by a gluon fusion process with the emission of an additional real gluon in the final state because of the Landau-Yang theorem,
\begin{equation}
g(p_a)\,{+}\,g(p_b)\,\to\, Q \overline Q  [^3S^{(1)}_1] (p_{\cal Q})\,{+}\,g(p_g) \, ,
\end{equation}
as described in detail in Appendix~\ref{app:unp-J}.
In the rest frame of the bound state, the relative momentum of the two quarks is small compared to their mass $m_Q$, which justifies a nonrelativistic approach. In agreement with Eq.~(A12) of Ref.~\cite{Baier:1983va}, we find that the corresponding partonic cross section can be written as
\begin{equation}
\frac{\d\hat \sigma}{\d\hat t} = \frac{\pi\alpha_s^3}{\hat s^2}\,H^U_{gg \to J/\psi g} \,,
\label{eq:unp-cs}
\end{equation}
with
\begin{equation}
H^U_{gg\to J/ \psi g}  = \frac{5}{9}\, \vert R_0(0)\vert^2 \, M\,
\frac{\hat s^2 (\hat s-M^2)^2 + \hat t^2 (\hat t-M^2)^2 + \hat u^2 (\hat u -M^2)^2 }
{(\hat s -M^2)^2 (\hat t -M^2)^2 (\hat u -M^2)^2}\,,
\label{eq:HU}
\end{equation}
where $R_0(0)$ is the value of the bound-state radial wave function at the origin. Details of the derivation are presented in Appendix~\ref{app:unp-J}. In the GPM approach, therefore, the unpolarized cross section for the process under study reads
\begin{align}
\d\sigma \equiv E_{\cal Q}\,\frac{\d\sigma}{\d^3\bfp_{\cal Q}} = & \frac{\alpha_s^3}{s} \int \frac{\d x_a}{x_a} \, \frac{\d x_b }{x_b} \; \d^2\bfk_{\perp a} \, \d^2\bfk_{\perp b} \,
     f_{g/p}(x_a,  k_{\perp a})
\> f_{g/p}(x_b, k_{\perp b})\, H_{gg\to J/\psi g}^U(\hat s, \hat t, \hat u) \> \delta(\hat s + \hat t + \hat u -M^2)\,,
\label{eq:den}
\end{align}
with $f_{g/p}(x, k_\perp)$ denoting the distribution of unpolarized gluons with light-front momentum fraction $x$ and transverse momentum $k_\perp = \vert \bm k_{\perp}\vert $. Its dependence on the hard scale of the process is not shown explicitly.

Concerning the $k_{\perp}$ dependence of the unpolarized gluon distributions, we use a simple factorized Gaussian parametrization
\begin{equation}
 f_{g/p}(x, k_\perp) = f_{g/p}(x) \, \frac{1}{\pi \langle k_\perp^2 \rangle} \,
e^{-k_\perp ^2/\langle k_\perp^2 \rangle} \>,
\label{eq:unp-TMD}
\end{equation}
with $\langle k_\perp^2 \rangle$ = 1 GeV$^2$ and $f_{g/p}(x)$ being the unpolarized gluon distribution, integrated over $k_\perp$, evaluated at the hard scale $M_T = \sqrt{\bm p_{T}^2 +M^2 }$, where $\bm p_T \equiv \bm p_{{\cal Q}T}$  is the transverse momentum of the $J/\psi$. Notice that the value adopted for the Gaussian width, for which no phenomenological information is currently available, has been fixed to optimize the description of $J/\psi$ data, within the uncertainties, in the low $p_T$ region.

\begin{figure}[t]
\begin{center}
\includegraphics[trim =  110 60 250 520,clip,width = 6.5cm]{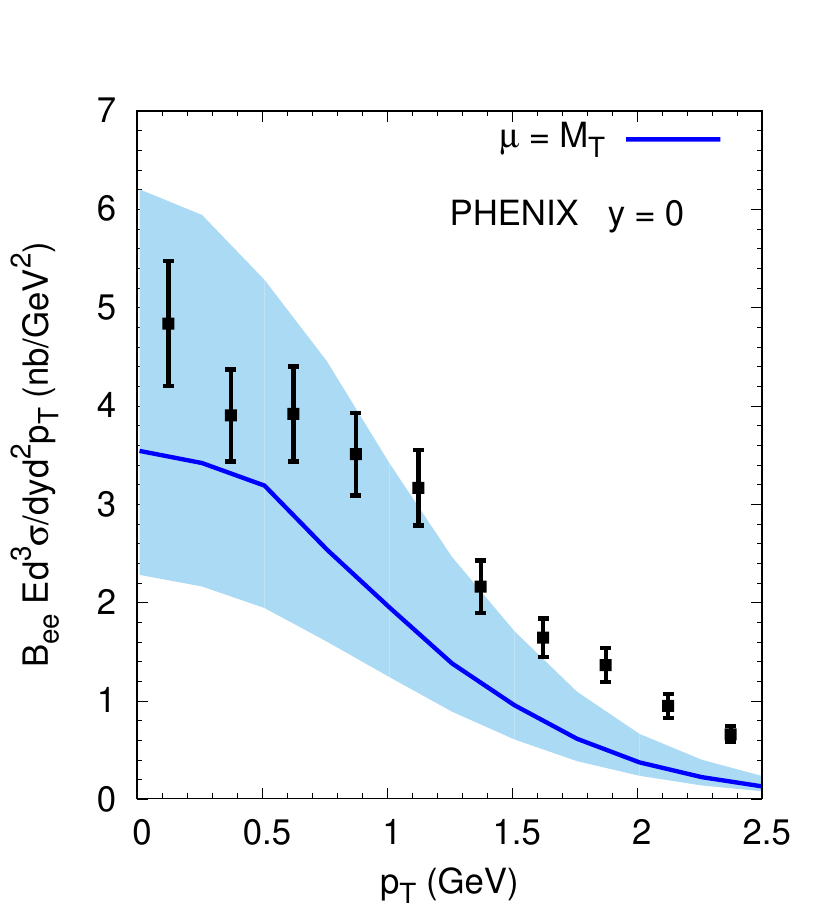}
\end{center}
\caption{Unpolarized cross section for the process $pp\to J/\psi \,X \to e^+e^-\,X$, at $\sqrt s$ = 200 GeV in the central rapidity region $y=0$, as a function of the transverse momentum $p_T$ of the $J/\psi$. The theoretical curve is obtained by  adopting the generalized parton model and the color-singlet production mechanism for the quarkonium. Data are taken from Ref.~\cite{Adare:2009js}. The uncertainty band results from varying  the factorization scale in the range $M_T/2 \le \mu \le 2 M_T$.}
\label{fig:unp-cs}
\end{figure}

In Fig.~\ref{fig:unp-cs} we compare our results for $J/\psi$ production, computed at rapidity $y=0$, with RHIC data taken at $\sqrt{s}= 200$ GeV and $\vert y \vert <  0.35$ from the PHENIX Collaboration~\cite{Adare:2009js}. We do not consider the analogous data from the STAR Collaboration~\cite{Adamczyk:2012ey,Adamczyk:2016dhc} since they cover mainly the region at larger $p_T$. For the parameters entering the cross section, we take $\vert R_0(0)\vert^2 = 1.01$ GeV$^3$, ${\rm Br}(J/\psi \to e^+e^-) = 0.0597$ and $M=3.097$ GeV.

These and the following results are based on the CTEQ6-LO parametrization of $f_{g/p}(x)$~\cite{Pumplin:2002vw}. The uncertainty band in the figure is obtained by varying  the factorization scale in the range $M_T/2 \le \mu \le 2 M_T$.  Since the data include not only the direct $J/\psi$ yield, but also feed-down contributions from $B$, $\psi (2S)$ and $\chi_c$ decays, our theoretical curves are divided by a factor of 0.58, which is the expected fraction of direct $J/\psi$ production~\cite{Lansberg:2010vq,Brodsky:2009cf}.
It turns out that, within the GPM approach and assuming a color-singlet production mechanism,  it is possible to reproduce RHIC data on $J/\psi$ cross sections reasonably well at small values of $p_T$, $p_T \le 2$ GeV.  This is in agreement with the findings of Refs.~\cite{Lansberg:2010vq,Brodsky:2009cf}, where
it is also shown how next-to-leading order QCD corrections and the contributions from the intrinsic charm
of the proton can further improve the theoretical description. The r\^ole of color-octet states~\cite{Bodwin:1994jh,Cho:1995vh,Cho:1995ce}, which becomes relevant at high $p_T$, seems to be much less important in the kinematic region under study. In our view, this justifies, at the present level of accuracy, the use of the CSM in our analysis of the PHENIX SSA data for $p^\uparrow p \to J/\psi\,X $  presented in the next section.

\subsection{Single-spin asymmetries in the GPM and CGI-GPM frameworks}
\label{SSAjpsi}

The SSA for the process $p^\uparrow p \to h\,X$ is defined by
\begin{equation}
A_N  \,\equiv \, \frac{\d \sigma^\uparrow - \d \sigma^\downarrow}{\d \sigma^\uparrow + \d \sigma^\downarrow}\, \equiv\, \frac{ \d\Delta\sigma}{ 2 \d\sigma}\,,
\label{eq:AN}
\end{equation}
where $\d \sigma^{\uparrow (\downarrow)}$ is the cross section for one of the initial nucleons polarized along the transverse direction $\uparrow  (\downarrow)$ with respect to the production plane. If we denote by  $ \hat f_{a/\pup}\,(x_a, \bfk_{\perp a}) $ the number density in momentum space of a parton $a$ inside a transversely polarized proton with mass $M_p$, the numerator of the asymmetry will be sensitive to
the difference~\cite{Bacchetta:2004jz}
\begin{align}
\Delta \hat f_{a/\pup}\,(x_a, \bfk_{\perp a})  \,&\equiv \,
\hat f_{a/\pup}\,(x_a, \bfk_{\perp a}) - \hat f_{a/p^\downarrow}\,
(x_a, \bfk_{\perp a})\nonumber \\
\label{defsiv}
&= \,\Delta^N f_{a/\pup}\,(x_a, k_{\perp a}) \> \cos\phi_a\nonumber \\
&=  \,-2 \, \frac{k_{\perp a}}{M_p} \, f_{1T}^{\perp a} (x_a, k_{\perp a}) \>
\cos\phi_a \,,
\end{align}
where $\Delta^N f_{a/\pup}(x_a, k_{\perp a})$ (or $f_{1T}^{\perp a} (x_a, k_{\perp a})$) is the Sivers distribution function for parton $a$ and $\phi_a$ is the azimuthal angle of its intrinsic transverse momentum $\bfk_{\perp a}$. The Sivers function satisfies the positivity bound
\begin{equation}
\vert \Delta^N f_{a/\pup}\,(x_a, k_{\perp a}) \vert  \le 2\,f_{a/p}\,(x_a, k_{\perp a})\,,~~{\rm or~equivalently}~~
\frac{k_\perp}{M_p}\, \vert f_{1T}^{\perp a} (x_a, k_{\perp a})\vert \le  f_{a/p}\,(x_a, k_{\perp a})~.
\label{eq:posbound}
\end{equation}
A more stringent constraint on the Sivers functions is given by the Burkardt sum rule (BSR)~\cite{Burkardt:2004ur}, which states that the total transverse momentum of all unpolarized partons inside a transversely polarized proton vanishes. Since available fits to the Sivers asymmetry for SIDIS data~\cite{Anselmino:2005ea,Anselmino:2008sga} almost fulfill, within uncertainties, the BSR, little room seems to be left for a gluon contribution. This is consistent with arguments valid in the large-$N_c$ limit of QCD~\cite{Efremov:2004tp}, according to which the gluon Sivers function should be suppressed by a factor $1/N_c$  as compared to the valence quark Sivers distributions at not-too-small values of $x$, namely $x\sim 1/N_c$.

Along the lines of Ref.~\cite{Anselmino:2005sh}, one finds that the numerator of the SSA for $J/\psi$ production in the GPM framework is given by
\begin{align}
 \d\Delta\sigma^{\rm GPM}\, \equiv \,& \frac{E_{\cal Q}\, \d\sigma^\uparrow}{\d^3\bfp_{\cal Q}} -
\frac{E_{\cal Q}\, \d\sigma^\downarrow}{\d^3\bfp_{\cal Q}} =  \frac{2\alpha_s^3}{s} \int \frac{\d x_a}{x_a} \, \frac{\d x_b }{x_b} \; \d^2\bfk_{\perp a} \, \d^2\bfk_{\perp b} \,
\nonumber \\
  &\times  \left  (- \frac{k_{\perp\,a}}{M_p}\right )  f_{1T}^{\perp\,g}(x_a,  k_{\perp a}) \cos\phi_a
\> f_{g/p}(x_b, k_{\perp b})\, H_{gg\to J/\psi g}^U(\hat s, \hat t, \hat u) \> \delta(\hat s + \hat t + \hat u -M^2)\,,
\label{eq:num-SSA}
\end{align}
where  $f_{g/p}$  and   $f_{1T}^{\perp\,g}$ are considered to be universal. The hard partonic function $H^{U}_{gg\to J/\psi g}$ is given explicitly in Eq.~(\ref{eq:HU}), while the denominator of the asymmetry is twice  Eq.~(\ref{eq:den}).

\begin{figure}[t]
\begin{center}
\subfloat[]{\includegraphics[trim= 200 600 100 50,clip,width=6.cm]{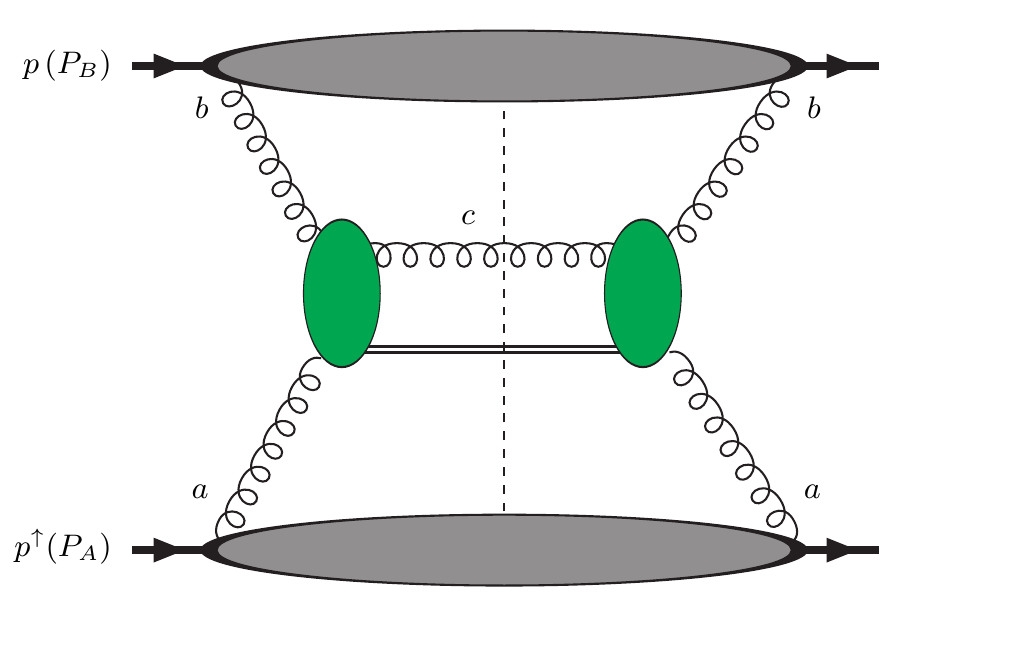}}
\subfloat[]{\includegraphics[trim= 200 600 100 50,clip,width=6.cm]{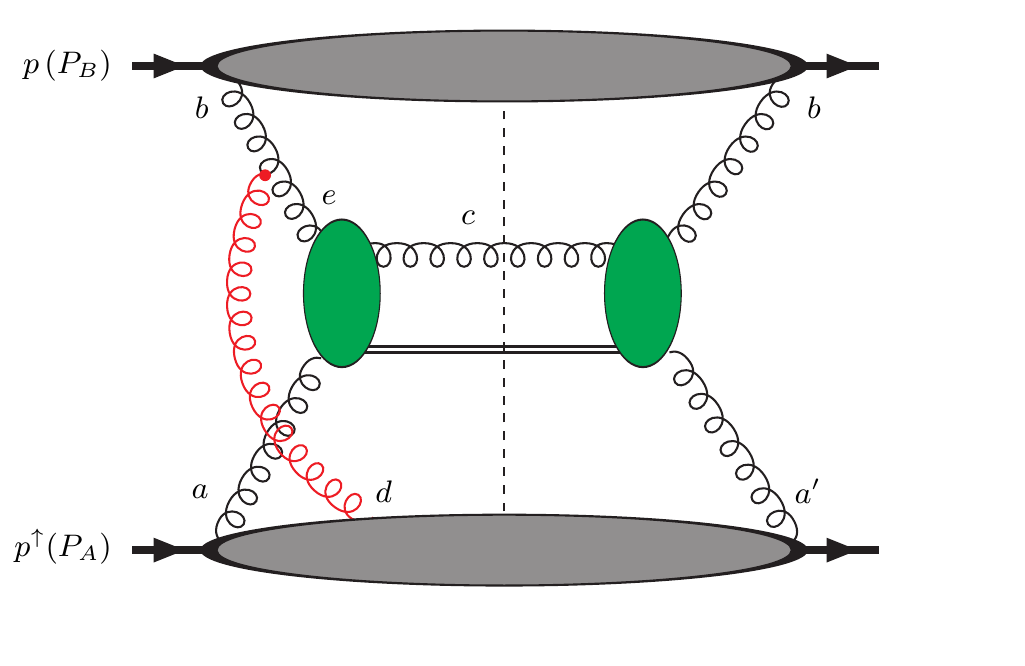}}
\subfloat[]{\includegraphics[trim= 200 600 100 50,clip,width=6.cm]{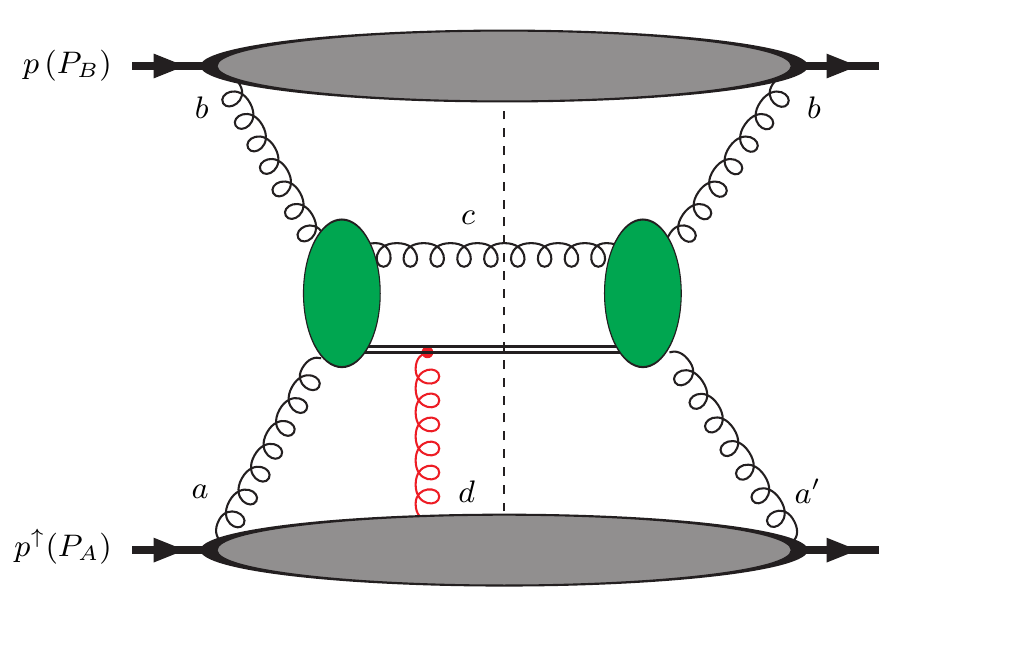}}
\end{center}
\caption{LO diagrams for the process $p^\uparrow p \to J/\psi\,X$ in the GPM formalism (a)
and in the CGI-GPM (b), in which the additional effect of initial-state interactions is included.  Final-state interactions (c) do not contribute when the $J/\psi$ is produced as a color singlet.
The scattering amplitudes for the underlying partonic reaction, $gg\to J/\psi\,g$, are represented by the central blobs, while the upper and lower ones describe the soft proton $\to$ gluon transitions.}
\label{fig:gglo-CGI}
\end{figure}

We now take into account the effects on the numerator of the asymmetry coming from initial- and final-state interactions between the struck parton (gluon) and the spectators from the polarized proton. Such interactions are encoded in the gauge links or Wilson lines that are needed in the definition of the Sivers function in terms of QCD operators to preserve gauge invariance, rendering it process dependent. In the framework of the CGI-GPM, ISIs and FSIs are approximated by a single, eikonal gluon, that corresponds to the leading-order contribution of the Wilson line in an expansion in the coupling constant $g_s$. It is therefore the imaginary part of the eikonal propagator that provides the phase needed to generate the Sivers asymmetry. Moreover, in the CGI-GPM, it is possible to express the process-dependent gluon Sivers function in Eq.~(\ref{eq:num-SSA}) as a linear combination of two independent and universal gluon distributions, denoted by $f_{1T}^{\perp\,g\,(f)}$ and $f_{1T}^{\perp\,g\,(d)}$, with coefficients that are calculable for each partonic process. The two distinct gluon Sivers distributions correspond to the two possible ways in which three gluon fields, with color indices $a$, $b$, $c$, can be neutralized, {\it i.e.\ } by contracting with either the antisymmetric ($T^a_{bc}\equiv -if_{abc}$) or the symmetric ($D^a_{bc}\equiv d_{abc}$) structure constants of the $SU(3)$-color group. Their first transverse moments,
\begin{equation}
f_{1 T}^{\perp\,g\,(f/d)}(x) = \int \d^2 \bm{k}_\perp\,\frac{k_\perp^2}{2M_p^2}\,f_{1T}^{\perp\,g\,(f/d)}(x,k_\perp)\,,
\end{equation}
at least at tree level are related to the two distinct trigluon Qiu-Sterman functions $T_G^{(f/d)}$, which have opposite behavior under charge conjugation. Hence $f_{1T}^{\perp\,g\,(f)}$ and $f_{1T}^{\perp\,g\,(d)}$ have different properties as well: for instance the former is $C$-even and expected to vanish in the small-$x$ region, whereas the latter is $C$-odd and not necessarily suppressed when $x$ is small~\cite{Schafer:2013opa}. Furthermore, only $f_{1T}^{\perp\,g\,(f)}$ is constrained by the BSR~\cite{Boer:2015vso}.

Formally, the numerator of the asymmetry in the CGI-GPM approach can be obtained from Eq.~(\ref{eq:num-SSA}) with the substitution
\begin{align}
f_{1T}^{\perp\,g}\, H_{gg\to J/\psi g}^U   \longrightarrow &~ \frac{C_I^{(f)} + C^{(f)}_{F_c}}{C_U}\, f_{1T}^{\perp\,g\,(f)}H_{gg\to J/\psi g}^U\,  + \, \frac{C_I^{(d)} + C^{(d)}_{F_c}}{C_U}\, f_{1T}^{\perp\,g\,(d)}H_{gg\to J/\psi g}^U\nonumber \\
 \equiv& ~ f_{1T}^{\perp\,g\,(f)}H_{gg\to J/\psi g}^{{\rm Inc} \,(f)}\,  + \, f_{1T}^{\perp\,g\,(d)}H_{gg\to J/\psi g}^{{\rm Inc}\,(d)}\,,
\end{align}
where we have introduced the modified partonic hard functions
\begin{align}
H_{gg\to J/\psi g}^{{\rm Inc} \,(f/d)} \equiv  \frac{C_I^{{\rm Inc}\,(f/d)}}{C_U}\,  H_{gg\to J/\psi g}^U =   \frac{C_I^{(f/d)} + C^{(f/d)}_{F_c}}{C_U}\,  H_{gg\to J/\psi g}^U~.
\label{eq:mod-H}
\end{align}
We have denoted with $C_{U}$ the color factor for the unpolarized cross section, which can be calculated from Fig.~\ref{fig:gglo-CGI}(a) knowing that the color factor for the scattering amplitude for $gg\to J/\psi\, g$ is $D^a_{bc}/2\sqrt{N_c}$, see Eq.~(\ref{eq:CFJ}). We find:
\begin{equation}
C_U= \frac{1}{(N_c^2-1)^2}\, \frac{1}{4 N_c}\, D^a_{bc}\, D^a_{cb} = \frac{1}{(N_c^2-1)^2}\, \frac{1}{4 N_c}\, \left [ \frac{(N_c^2-4)(N_c^2-1)}{N_c} \right ] = \frac{N_c^2-4}{4 N_c^2(N_c^2-1)} = \frac{5}{288}\,,
\label{eq:CU}
\end{equation}
where we have substituted $N_c=3$ in the last equality. In order to compute the new color factors $C_I^{(f/d)}$ and $C_{F_c}^{(f/d)}$ for the ISIs and FSIs respectively, in the following we will adopt the methods developed for the twist-three, three-gluon correlation functions~\cite{Ji:1992eu,Kang:2008qh,Kang:2008ih}.

We consider first the effects of ISIs, described by the insertion of a longitudinally polarized gluon $A^+$ with momentum $k^\mu\approx k^+$ and color index $d$, as depicted in Fig.~\ref{fig:gglo-CGI}(b). The corresponding amplitude squared can be obtained from the Born one in Fig.~\ref{fig:gglo-CGI}(a) with the replacement 
\begin{align}
\varepsilon^\nu_{\lambda_b}(p_b)   \longrightarrow  & \, \varepsilon^\nu_{\lambda_b}(p_b)\, A^{\rho}(k)\,(-g_s\,f_{bed})\, \left [ (k-p_b)_\lambda \,g_{\nu\rho} + (2 p_b + k)_\rho \,g_{\nu \lambda} - (p_b+2 k)_\nu \,g_{\rho \lambda} \right ]\, \frac{-i g^{\lambda\sigma}}{(k+p_b)^2+i\epsilon}\nonumber \\
 & \approx \, \varepsilon^\nu_{\lambda_b}(p_b)\, A^{\rho}(k)\,(-i g_s\,T^d_{be})\,( -p_{b}^{\sigma}\, g_{\nu\rho}+ 2 p_{b\rho} \,g_{\nu}^{\sigma} - p_{b\nu}\,g_{\rho}^{\sigma})\, \frac{-i}{2k^+p_b^-+i\epsilon}\nonumber \\
 & \approx \varepsilon^\sigma_{\lambda_b}(p_b)\,A^{+}(k)\,(g_s\,T^d_{eb}) \, 2 p_{b}^-\,\frac{1}{2k^+p_b^-+i\epsilon}\nonumber \\
 & = \varepsilon^\sigma_{\lambda_b}(p_b)\,\left  [ {g_s A^+(k)}\, \frac{1}{k^+ + i \epsilon}\right ]\,  T^d_{eb} \,,
 \label{eq:ggg}
\end{align}
where, according to the eikonal approximation, in the numerator in the second line we have neglected all $k^\mu$ components with respect to the components of $p_b^\mu$.   Moreover, we choose the polarization vector of the external gluon such that $\varepsilon_{\lambda_b}^-(p_b)=0$. This,  together with the orthogonality condition $\varepsilon_\nu(p_b)  p_b^\nu=0$,  leads us to the final result in Eq.~(\ref{eq:ggg}). By using the relation
\begin{equation}
\frac{1}{k^+ \pm i\epsilon} = \text{P}\, \frac{1}{k^+}  \mp i \pi \delta(k^+)\,,
\end{equation}
where P denotes the principal value, we find that the imaginary part of the quark propagator, $1/(k^+ + i\epsilon)$, is given by  $-i\pi\delta(k^+)$. In the calculation of the full diagram, such term is multiplied by the Born amplitude, taken with a different color factor $T^d_{eb}$ because of the presence of an extra gluon. At this point we define the color projectors
\begin{equation}
{\cal T}^{c}_{a a^\prime}   =  {\cal N}_{\cal T}\, T^c_{a a^\prime}\,, \qquad \qquad {\cal D}^{c}_{a a^\prime}   =  {\cal N}_{\cal D}\, D^c_{a a^\prime}\, ,
\label{eq:cp}
\end{equation}
with
\begin{equation}
 {\cal N}_{\cal T} = \frac{1}{{\rm Tr}[T^cT^c]} = \frac{1}{N_c(N_c^2-1)}\,,\qquad \qquad  {\cal N}_{\cal D} = \frac{1}{{\rm Tr}[D^cD^c]} = \frac{N_c}{(N_c^2-4)(N_c^2-1)}\,,
 \label{eq:norm-cp}
\end{equation}
corresponding to the two different ways in which color can be neutralized. For the $f$-type gluon Sivers function, the relative color factor is therefore calculated from Fig.~\ref{fig:gglo-CGI}(b) as follows
\begin{align}
C_{I}^{(f)} & =\frac{1}{N_{c}^{2}-1}{\cal T}_{aa^{\prime}}^{d}T_{eb}^{d} \,\frac{1}{4 {N_c}}\, D^e_{ac} D^{b}_{ca'} \nonumber \\
& = - \frac{1}{4 N_c^2(N_c^2-1)^2}\,  T^d_{a^\prime a} D^{c}_{a e} T^d_{eb}D^c_{ba^{\prime}}\,,\nonumber \\
& =-\frac{1}{2}C_{U},\label{eq:CI}
\end{align}
where we have used the identity  ${\rm Tr}\left [ T^dD^cT^dD^c \right]  =  (N_c^2-1) (N_c^2-4)/2$.  Likewise, for the $d$-type color factor, we find
\begin{align}
C_{I}^{(d)} & = \frac{1}{N_{c}^{2}-1}{\cal D}_{aa^{\prime}}^{d}T_{eb}^{d} \,\frac{1}{ 4 {N_c}}\, D^e_{ac} D^{b}_{ca'} = 0\,,
\end{align}
since ${\rm Tr}\left [ D^dD^cT^dD^c \right]  = 0$. As already pointed out in Ref.~\cite{Yuan:2008vn}, the net contribution of the heavy quark-antiquark pair to the FSI, depicted in Fig.~\ref{fig:gglo-CGI}(c), is zero because the pair is produced in a color-singlet state.  Hence we have
\begin{align}
C_{F_c}^{(f)} & =  C_{F_c}^{(d)} = 0
\end{align}
and therefore
\begin{equation}
H^{{\rm Inc} \,(f)}_{gg\to J/\psi g}  = -\frac{1}{2}\, H^{U}_{gg\to J/\psi g}\,, \qquad H^{{\rm Inc} \,(d)}_{gg\to J/\psi g}  =0.
\end{equation}
We note that we did not consider the FSIs of the unobserved particle (gluon) because they are known to vanish after summing the different cut diagrams, see for example the discussion in Ref.~\cite{Gamberg:2010tj}.

We then find that $f_{1T}^{\perp\,g\,(d)}$ does not contribute to the specific partonic reaction under study, $gg\to J/\psi\,g$, and that the numerator of the SSA can be expressed as a convolution of $f_{1T}^{\perp\,g\,(f)}$ with a modified partonic hard function $ H_{gg\to J/\psi g}^{{\rm Inc}\,(f)} $  as follows
\begin{align}
 \d\Delta\sigma^{\rm CGI}\, \equiv \,& \frac{E_{\cal Q}\, \d\sigma^\uparrow}{\d^3\bfp_{\cal Q}} -
\frac{E_{\cal Q}\, \d\sigma^\downarrow}{\d^3\bfp_{\cal Q}} =  \frac{2\alpha_s^3}{s} \int \frac{\d x_a}{x_a} \, \frac{\d x_b }{x_b} \; \d^2\bfk_{\perp a} \, \d^2\bfk_{\perp b} \,
\nonumber \\
 &\times  \left  (- \frac{k_{\perp\,a}}{M_p}\right )  f_{1T}^{\perp\,g\,(f)}(x_a,  k_{\perp a}) \cos\phi_a \> f_{g/p}(x_b, k_{\perp b})\, H_{gg\to J/\psi g}^{{\rm Inc}\,(f)}(\hat s, \hat t, \hat u) \>  \delta (\hat s + \hat t + \hat u  -M^2 )~.
\end{align}
This process can therefore be very useful to gather separate and direct information on $f_{1T}^{\perp\,g\,(f)}$.

\subsection{Numerical results}
\label{Jpsi-res}

A first extraction of the gluon Sivers distribution~\cite{D'Alesio:2015uta}, obtained by fitting very precise, RHIC midrapidity data on $A_N$ for inclusive neutral pion production~\cite{Adare:2013ekj} within a GPM approach, showed that $f_{1T}^{\perp\,g}$ is very small with respect to its theoretical positivity bound in Eq.~(\ref{eq:posbound}). In that analysis the following functional form for $\Delta^N f_{g/\pup}$ was adopted:
\begin{equation}
\Delta^N\! f_{g/p^\uparrow}(x,k_\perp) =   \left (-2\frac{k_\perp}{M_p}  \right )f_{1T}^{\perp\,g} (x,k_\perp)  = 2 \, {\cal N}_g(x)\,f_{g/p}(x)\,
h(k_\perp)\,\frac{e^{-k_\perp^2/\langle k_\perp^2 \rangle}}
{\pi \langle k_\perp^2 \rangle}\,,
\label{eq:siv-par-1}
\end{equation}
where
\begin{equation}
{\cal N}_g(x) = N_g x^{\alpha}(1-x)^{\beta}\,
\frac{(\alpha+\beta)^{(\alpha+\beta)}}
{\alpha^{\alpha}\beta^{\beta}}\,,
\label{eq:nq-coll}
\end{equation}
with $|N_g|\leq 1$ and
\begin{equation}
h(k_\perp) = \sqrt{2e}\,\frac{k_\perp}{M'}\,e^{-k_\perp^2/M'^2}\,.
\label{eq:h-siv}
\end{equation}
With the above choice the Sivers function automatically fulfills its positivity bound for any $(x,k_\perp)$ values. Alternatively, if we define the parameter
\begin{equation}
\rho = \frac{M'^2}{\langle k_\perp^2 \rangle +M'^2}\, ,
\label{eq:rho}
\end{equation}
such that $0< \rho < 1$, then  Eq.~(\ref{eq:siv-par-1}) becomes
\begin{equation}
\Delta^N\! f_{g/p^\uparrow}(x,k_\perp) =   2 \,  \frac{\sqrt{2e}}{\pi}   \, {\cal N}_g(x)\,f_{g/p}(x)\,\sqrt{\frac{1-\rho}{\rho}}\,k_\perp\,
\frac{e^{-k_\perp^2/ \rho\langle k_\perp^2 \rangle}}
{\langle k_\perp^2 \rangle^{3/2}}~.
\label{eq:siv-par}
\end{equation}
In Ref.~\cite{D'Alesio:2015uta} the value of $\langle k_\perp^2\rangle$ was taken to be $ \langle k_\perp^2\rangle = 0.25$ GeV$^2$, while the parameters $N_g$, $\alpha$, $\beta$, $\rho$ were fitted to the data.

Here, as already stated in the Introduction, we do not use any information from the previous analysis, and start adopting the value $\langle k_\perp^2\rangle = 1$ GeV$^2$, according to the results shown in Sec.~\ref{csm}. On the other hand, in order to maximize the effect, we saturate the positivity bound for the $x$-dependent part (i.e.~we take ${\cal N}_g(x)=\pm 1$) and adopt the value $\rho = 2/3$~\cite{D'Alesio:2010am} in Eq.~(\ref{eq:siv-par}). For the unpolarized gluon distribution $f_{g/p}(x)$ we use the CTEQ6-LO parametrization as before, with the factorization scale equal to $M_T$.

\begin{figure}[t]
\begin{center}
\includegraphics[trim= 50 60 190 500,clip, width=8.7cm]{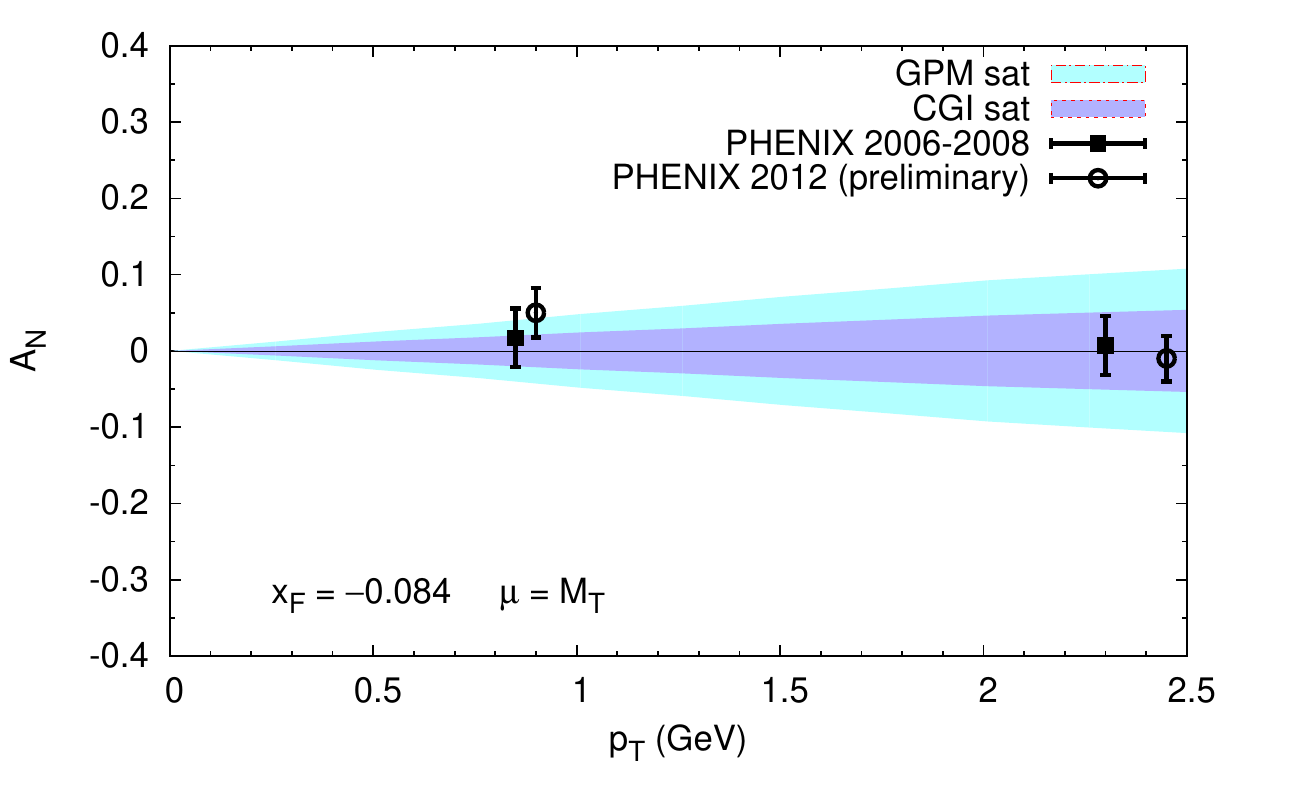}
\includegraphics[trim= 50 60 190 500,clip,width=8.7cm]{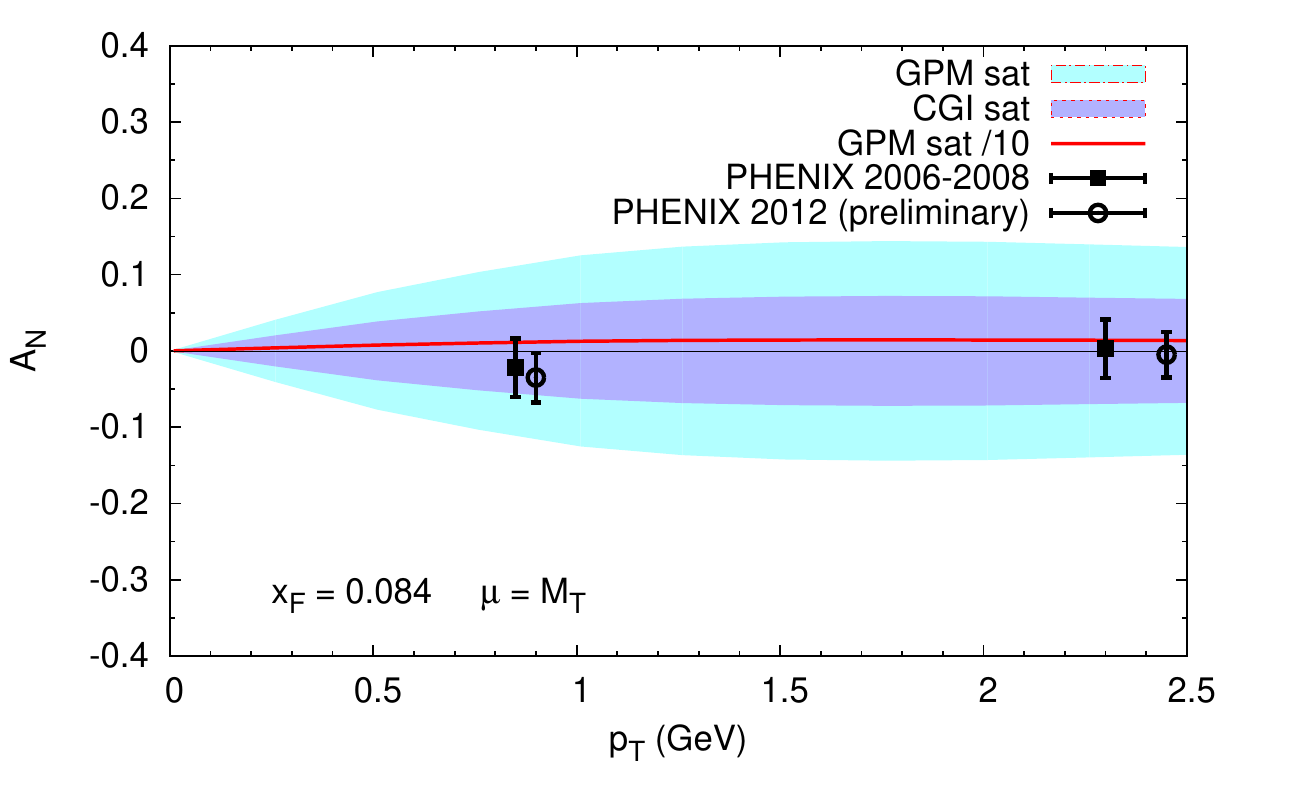}

\includegraphics[trim= 50 60 190 500 ,clip, width=8.7cm]{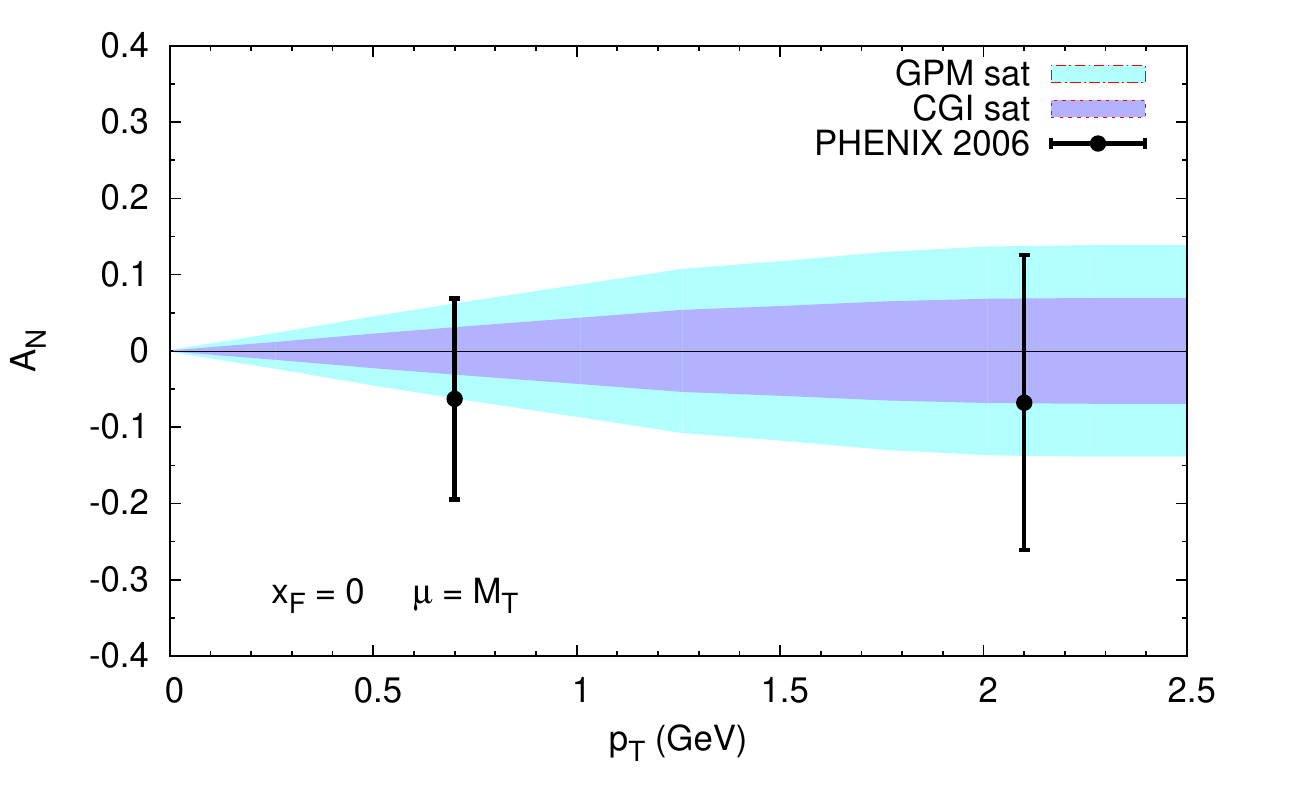}
\includegraphics[trim= 50 60 190 500,clip,width=8.7cm]{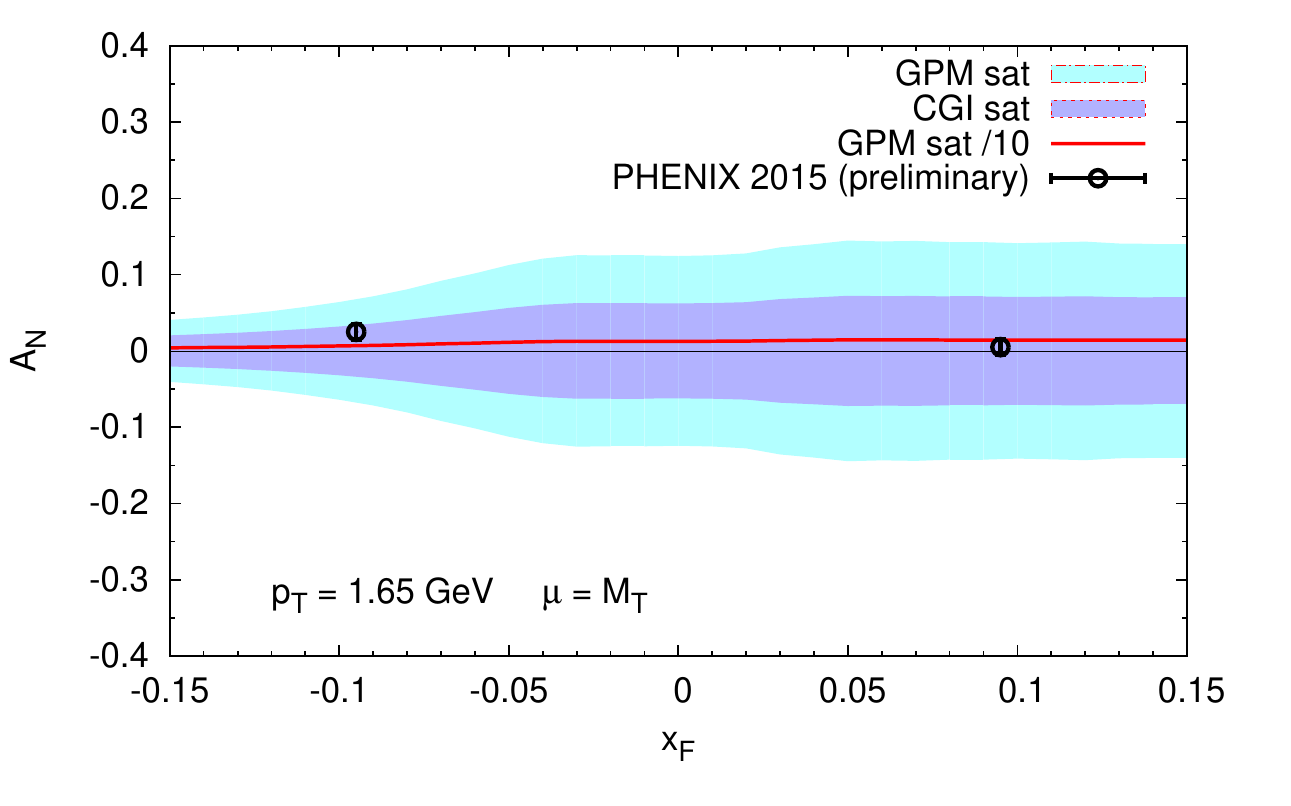}
\caption{Comparison of the available data from PHENIX~\cite{Adare:2010bd,ChenXu:spin2016} with the bands of possible values of $A_N$, between the lower and the upper bounds (${\cal N}_{g}(x)=\pm 1$), for the process $p^\uparrow p \to J/\psi \,X $ at $\sqrt s = 200$ GeV, calculated in both the GPM and CGI-GPM approaches. Upper panels: as a function of $p_T$ at $x_F=-0.084$ (left) and $x_F=+0.084$ (right). Lower panels: as a function of $p_T$ at $x_F=0$ (left) and as a function of $x_F$ at $p_T=1.65$ GeV (right). The red solid lines represent an estimate obtained with ${\cal N}_g(x)= + 0.1$ within the GPM approach (see text for details).}
\label{fig:AN-J}
\end{center}
\end{figure}

Our results for the bands of possible values of $A_N$, between the lower and the upper bounds (${\cal N}_g(x)=\pm 1$), calculated both in the GPM and the CGI-GPM at $\sqrt s = 200$ GeV, are confronted in Fig.~\ref{fig:AN-J} with PHENIX data~\cite{Adare:2010bd,ChenXu:spin2016}.

As expected from the theoretical calculation, $A_N$ in the CGI-GPM is a factor of 2 smaller (in size) as compared to the GPM prediction. We note that, since $A_N$ is the ratio of two cross sections, it is much less sensitive to the choice of the factorization scale than the unpolarized cross section presented in Fig.~\ref{fig:unp-cs}.

It turns out that the 2006 data at $x_F=0$ (lower-left panel) are not able to give any constraint or discriminate among the two models. Only the combined 2006-2008 and the preliminary 2012 data at $x_F=0.084$ (upper-right panel), and partially also at $x_F=-0.084$ (upper-left panel), are precise enough to further constrain the magnitude of the gluon Sivers function within the GPM approach. As an example, assuming the validity of the GPM, in the upper-right panel of Fig.~\ref{fig:AN-J} the red solid curve illustrates how a (positive) Sivers distribution reduced by one order of magnitude w.r.t.~its positivity bound would be in better agreement with the measurements. The latest preliminary data (RUN 2015) at fixed $p_T=1.65$ GeV (lower-right panel) are even more important since, thanks to their high accuracy, they could constrain the GSF, not only within the GPM, but also in the CGI-GPM approach. Once again the red solid line represents an estimate within the GPM obtained adopting ${\cal N}_g(x)=+0.1$.

On the other hand, the overall present precision as well as the amount of the data does not allow us to reject any of the two models. To this end, it would be helpful to determine the sign of  $f_{1T}^{\perp\,g\,(f)}$ independently, for example from a study of the process $p^\uparrow p \to D\,X$, as described in the next section.

\section{Single-spin asymmetries in $p^\uparrow p \to D\,X$}
\label{SSA-D}

\subsection{GPM and CGI-GPM formalism}
\label{Dmes-form}

We now turn to the study of the process
\begin{equation}
p^\uparrow(p_A)\,{+}\,p(p_B)\,\to\, D (p_{D}) \, {+}  \,X \, ,
\label{eq:proc-D}
\end{equation}
which has been already analyzed within both the GPM~\cite{Anselmino:2004nk,Godbole:2016tvq} and the twist-three frameworks~\cite{Kang:2008ih,Koike:2011mb}. As already discussed in Ref.~\cite{Anselmino:2004nk}, to which we refer for details, $D$ mesons are produced from the fragmentation of a $c$ or $\bar c$ quark created either through annihilation of a light quark pair, $q\bar q \to c\bar c$, or through gluon fusion, $gg\to c\bar c$. The unpolarized cross section can therefore be written as
\begin{align}
2 \d\sigma  \equiv
\frac{E_D \, \d\sigma^{\uparrow}} {\d^{3} \bfp_D} +
\frac{E_D \, \d\sigma^{\downarrow }} {\d^{3} \bfp_D}   \,= &\, \frac{2\alpha_s^2}{s}
\int \frac{\d x_a}{x_a} \, \frac{\d x_b}{x_b} \, \d z\, \d^2 \bfk_{\perp a} \, \d^2 \bfk_{\perp b} \,
\d^3 \bfk_D \,
\delta (\bfk_D \cdot \hat{\bfp}_Q) \,
\delta (\hat s +\hat t +\hat u - 2m_c^2) \nonumber \\
& ~~\times \,  {\mathcal J}(z,\bfk_D) \Biggl\{ \sum_q
\left[  f_{q/p}(x_a, k_{\perp a}) \>
 f_{\bar q/p}(x_b,  k_{\perp b}) \>
H^U_{q \bar q \to Q \overline Q}(\hat s, \hat t, \hat u) \, D_{D/Q}(z,\bfk_D) \>
 \right] \nonumber \\
&  ~~\hspace{1cm}+
\left[ f_{g/p}(x_a, k_{\perp a}) \> f_{g/p}(x_b,  k_{\perp b}) \>
H^U_{gg \to Q \overline Q} (\hat s, \hat t , \hat u) \>
 D_{D/Q}(z,\bfk_D) \right] \Biggr\} \,,
\end{align}
where $z$ is the light-cone momentum fraction of the parton $Q$ carried by the $D$ meson, $m_c$ is the (anti)charm mass, $q = u, \bar u, d, \bar d, s, \bar s$, and $Q = c$  if $D= D^+, D^0$ or $Q = \bar c$ if $D= D^-, \overline D^0$. We choose the reference frame such that the polarized proton moves along the $Z$ axis, with polarization $\uparrow$ along the positive $Y$ axis, and $XZ$  is the production plane. This means that $\bm k_{\perp a}$ and $\bm k_{\perp b} $ have only $X$ and $Y$ components, while $\bm k_D$ has a $Z$ component as well. The function $\delta ( \bm k_D \cdot \hat{\bm p}_Q) $ is hence  needed to perform the integral only over $\bm k_{\perp  D}$, {\it i.e.}\  the components of $\bm k_D$ which are transverse w.r.t.\ the direction of the fragmenting quark  $\hat{\bm p}_Q$. The Jacobian connecting the partonic to the observed hadronic phase space reads
\begin{equation}
{\mathcal J} (z,\bfk_D) = \frac{1}{ z^2}\,
\frac{ \big( E_D+\sqrt{\bfp_D^2 - \bfk_{\perp D}^2} \big)^2}
{4(\bfp_D^2 - \bfk_{\perp D}^2)} \,
\Bigg[1- \frac{z^2 m_c^2}
{ \big( E_D+\sqrt{\bfp_D^2 - \bfk_{\perp D}^2} \big)^2}\Bigg]^2  \,.
\end{equation}
Moreover, the partonic cross sections are written in the form
\begin{equation}
\frac{\d \hat \sigma}{\d\hat t } = \frac{\pi\alpha_s^2}{\tilde s^2}\,H_{ab\to c d},
\end{equation}
with
\begin{align}
H_{q\bar q\to c\bar c}^U \, = \,&   \frac{N_c^2-1}{2 N_c^2}\,\left (  \frac{\tilde t^2 + \tilde u^2 + 2 m_c^2 \tilde s}{\tilde s^2}  \right )\, , \nonumber \\
H_{g g\to c\bar c}^U \, =  \,&    \frac{N_c}{N_c^2-1} \, \frac{1}{\tilde t \tilde u}\,\left ( \frac{N_c^2-1}{2N_c^2} - \frac{\tilde t \tilde u}{\tilde s^2}    \right ) \left ( \tilde t^2 + \tilde u^2 + 4 m_c^2 \tilde s - \frac{4 m_c^4 \tilde s^2}{\tilde t \tilde u }    \right ) \, ,
 \label{eq:unp-D}
\end{align}
where we have introduced the following invariants:
\begin{equation}
\tilde s  \equiv (p_a+p_b)^2 = \hat s\,,\qquad   \tilde{t} \equiv (p_a-p_c)^2-m_c^2= \hat t -m_c^2\,,\qquad \tilde{u} \equiv  (p_b-p_c)^2-m_c^2 = \hat u -m_c^2~.
\end{equation}

In the GPM approach the numerator of the asymmetry for the process under study reads~\cite{Anselmino:2004nk}
\begin{align}
\d \Delta \sigma^{\rm GPM} \,\equiv & \,
\frac{E_D \, \d\sigma^{\uparrow}} {\d^{3} \bfp_D} -
\frac{E_D \, \d\sigma^{\downarrow}} {\d^{3} \bfp_D}   = \frac{2\alpha_s^2}{s}
\int \frac{\d x_a}{x_a} \, \frac{\d x_b}{x_b}  \, \d z \, \d^2 \bfk_{\perp a} \, \d^2 \bfk_{\perp b} \,
\d^3 \bfk_{D} \,
\delta (\bfk_{D} \cdot \hat{\bfp}_c) \,
\delta (\hat s +\hat t +\hat u - 2m_c^2) \> \nonumber \\
& \times {\mathcal J}(z,\bfk_D) \, \Biggl\{ \sum_q
\left[ \left ( -\frac{k_{\perp\,a }}{M_p}  \right ) f_{1T}^{\perp\, q}(x_a, k_{\perp a})\cos\phi_a \,f_{\bar q/p}(x_b,
k_{\perp b}) \>
H^U_{q\bar q\to Q\overline Q} (\hat s, \hat t , \hat u)\,
D_{D/Q}(z,\bfk_D) \right]  \nonumber \\
& \, +
\left[  \left ( -\frac{k_{\perp\,a }}{M_p}  \right ) f_{1T}^{\perp\, g}(x_a, k_{\perp a})  \cos\phi_a\,  f_{g/p}(x_b,  k_{\perp b}) \>
H^U_{gg\to Q \overline Q}(\hat s, \hat t, \hat u)\,
 D_{D/Q}(z,\bfk_D) \right] \Biggr\}\,,
\label{eq:den-D}
\end{align}
with $H^U_{q \bar q\to Q \overline Q}$ and  $H^U_{gg\to Q \overline Q}$ given by Eqs.~(\ref{eq:unp-D}).

Notice that, as the gluons cannot carry any transverse spin, the elementary process $gg\to c\bar c$ results in unpolarized final quarks. In the $q\bar q\to c\bar c$ process one of the initial partons (the one inside the transversely polarized proton) can be polarized; however, there is no single-spin transfer in this $s$-channel interaction so that the final $c$ and $\bar c$ are again not polarized. Moreover, even when they are produced in the process $q^\uparrow \bar q^\uparrow  \to c \bar c$, where the initial quarks are transversely polarized because of the Boer-Mulders effect~\cite{Boer:1999mm}, the $s$-channel annihilation does not create a polarized final $c$ or $\bar c$. Consequently, there cannot be any Collins fragmentation contribution to $A_N$. More generally, it has been checked that all contributions to $A_N$, other than the Sivers one, enter with azimuthal phase factors that strongly suppress them after integration over transverse momenta. Hence they can be safely neglected~\cite{Anselmino:2004nk}.

\begin{figure}[t]
\begin{center}
\includegraphics[trim= 50 710 60 45,clip,width=17cm]{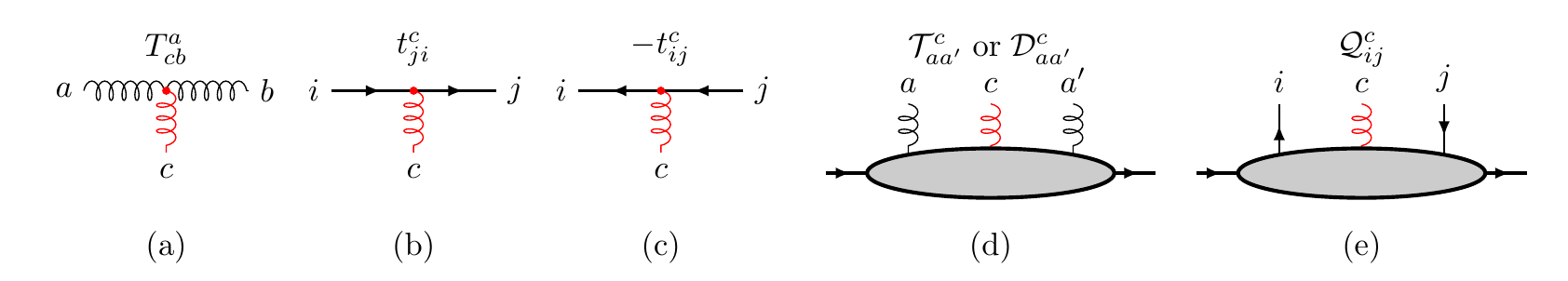}
\end{center}
\caption{CGI-GPM color rules for  the eikonal three-gluon (a),  quark-gluon (b) and  antiquark-gluon (c) vertices.  The
color projectors for the gluon (d) and the quark Sivers functions (e) are shown as well. The eikonal gluon has color index $c$.}
\label{fig:Feynman}
\end{figure}

In the CGI-GPM framework, the Sivers functions become process dependent because both ISIs and FSIs are taken into account.
Starting with the $q\bar q$ subprocess, in the calculation of the asymmetry one can still use the (anti)quark Sivers distributions extracted from SIDIS measurements, but they have to be convoluted with the following partonic hard functions
\begin{align}
H^{{\rm Inc}}_{q\bar q\to c\bar c} & = - H^{{\rm Inc}}_{\bar q q\to \bar c c} =  \frac{N_c^2-1}{2N_c^2}\,\left (  \frac{\tilde t^2 + \tilde u^2 + 2 m_c^2\tilde s}{\tilde s^2} \right )\, , \nonumber\\
H^{{\rm Inc}}_{q\bar q\to  \bar c  c} & = - H^{{\rm Inc}}_{\bar q q \to   c \bar c} = \frac{3}{2}\,\frac{1}{N_c^2}\,\left (  \frac{\tilde t^2 + \tilde u^2 + 2 m_c^2\tilde s}{\tilde s^2} \right )~.
\label{eq:Hq}
\end{align}
The relative color factors have been derived as described in the previous section and in Ref.~\cite{Gamberg:2010tj}, using the color assignments collected in Fig.~\ref{fig:Feynman}, by means of the color projectors of Eq.~(\ref{eq:cp}) and the additional one
 \begin{equation}
 {\cal Q}^c_{ij}= {\cal N}_{\cal Q} \, t^c_{ij} \,,
 \end{equation}
where $t^c_{ij}$  are the generators of $SU(N_c)$ in the fundamental representation and
 \begin{equation}
  {\cal N}_{\cal Q} = \frac{1}{ \text{Tr}[t^ct^c]} = \frac{2}{N_c^2-1}~.
 \end{equation}
We point out that Eqs.~(\ref{eq:Hq}) are in agreement with the twist-three expressions  in Ref.~\cite{Koike:2011mb} and, in the massless limit, with the CGI-GPM partonic functions in Ref.~\cite{Gamberg:2010tj}.

Turning to the gluon induced subprocess $gg\to c\bar c$, the effects of ISIs and FSIs have to be estimated diagram by diagram. The resulting color factors are presented in Table~\ref{tab:gg2ccb}. As in the previous section, $C_U$ denotes the usual unpolarized color factor for the specific diagram $D$, while $C_I^{(f/d)}$, $C_{F_c}^{(f/d)}$, $C_{F_d}^{(f/d)}$ are the color factors obtained when an extra gluon is attached in $D$ to parton $b$ (the gluon from the unpolarized proton), parton $c$ (the charm quark fragmenting into the observed $D$ meson, in this case) or parton $d$ (the unobserved anticharm quark, here) respectively. Once again, the two labels $f$ and $d$ distinguish between the two possible ways in which color is neutralized, leading to the two independent gluon Sivers functions. Furthermore,  $C^{\text{Inc}\,(f/d)} \equiv C_I^{(f/d)} + C^{(f/d)}_{F_c}$. A detailed derivation of these color factors for the first diagram in Table~\ref{tab:gg2ccb} is provided in Appendix~\ref{app:CGI-D} for illustration. Finally, we point out that our {\it gluonic pole strengths}, defined as
\begin{equation}
C_G^{(f/d)} \equiv \frac{C_I^{(f/d)} + C_{F_c}^{(f/d)} + C_{F_d}^{(f/d)}}{C_U} \,,
\end{equation}
are in full agreement with the ones given in Ref.~\cite{Bomhof:2006ra} for the study of the gluon Sivers effect in less inclusive processes like $p^\uparrow p \to \pi\,\pi\,X$, for which the FSIs of parton $d$ need to be taken into account as well. Notice that the results in Ref.~\cite{Bomhof:2006ra} have been derived adopting a different method, {\it i.e.}~by looking at the full gauge link structure and taking the derivative of the gauge link. We have checked that the one-gluon approximation employed here, which consists in considering only the first-order contribution of the gauge link in an expansion in terms of the strong coupling $g_s$, is sufficient to recover the exact gluonic pole strengths in any partonic process calculated at LO in perturbative QCD~\cite{Bomhof:2006ra,dalesio2017}.

\begin{table}[t]
\begin{center}
\includegraphics[trim= 50 480 0 90,clip,width=17cm]{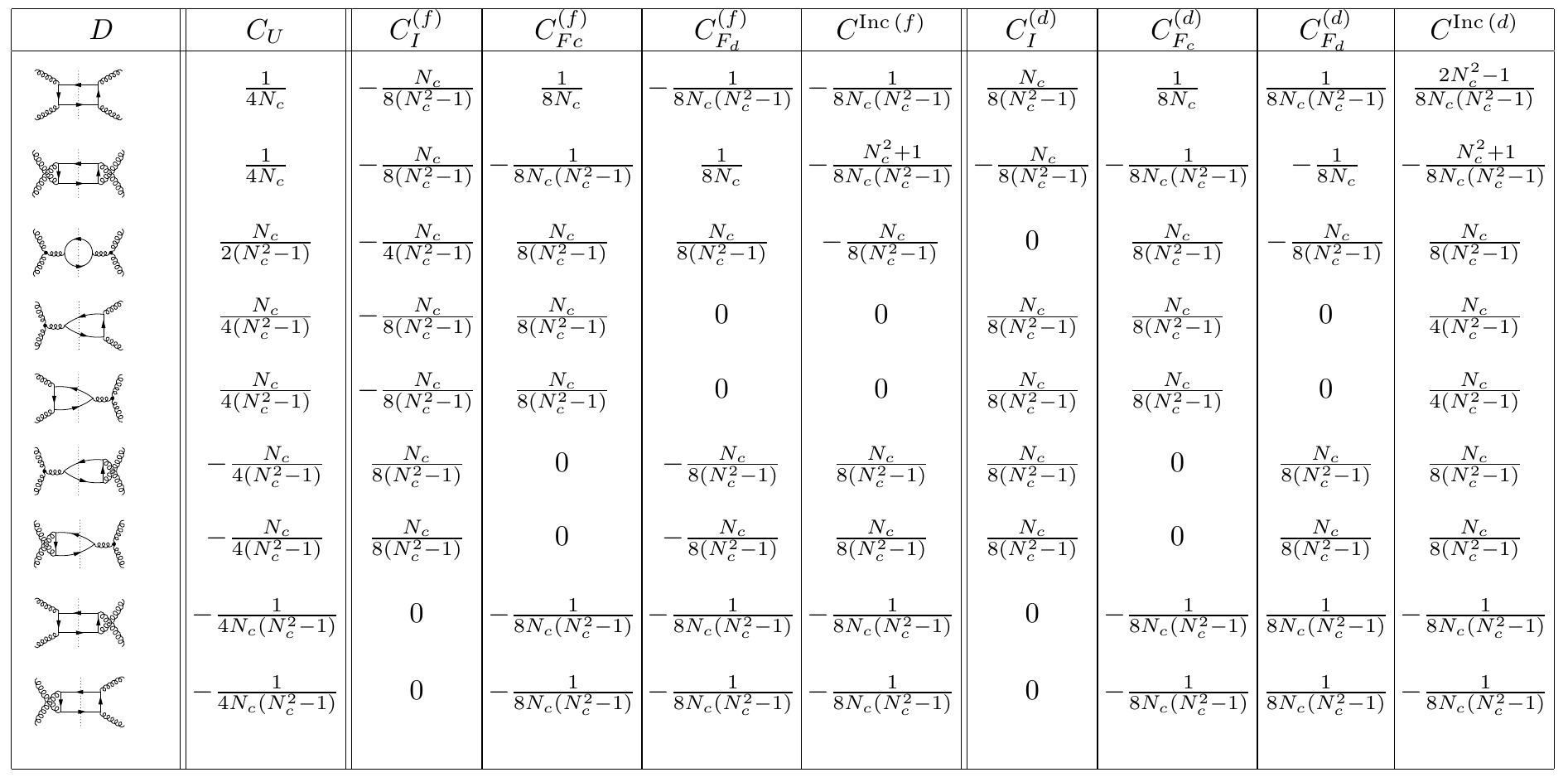}
\end{center}
\caption{Color factors for the LO diagrams contributing to the process $gg\to c\bar c$ in the CGI-GPM approach.}
\label{tab:gg2ccb}
\end{table}

By summing all the diagrams, taken with the new color factors $C^{\text{Inc}\,(f/d)}$, one obtains  $H^{\text{Inc}\,(f/d)}$, defined as in Eq.~(\ref{eq:mod-H}) for the subprocess $gg\to J/\psi\,g$.
At variance with what we have found for $p^\uparrow p \to J/\psi\,X$, both the independent gluon Sivers functions $f_{1T}^{\perp\,g\,(f)}$ and  $f_{1T}^{\perp\,g\,(d)}$ contribute to $A_N$ for  $p^\uparrow p \to D\,X$. Explicitly, the numerator of the asymmetry reads
\begin{align}
\d \Delta \sigma^{\rm CGI} \,\equiv & \,
\frac{E_D \, \d\sigma^{\uparrow}} {\d^{3} \bfp_D} -
\frac{E_D \, \d\sigma^{\downarrow}} {\d^{3} \bfp_D}   = \frac{2\alpha_s^2}{s}
\int \frac{\d x_a}{x_a} \, \frac{\d x_b}{x_b}  \, \d z \, \d^2 \bfk_{\perp a} \, \d^2 \bfk_{\perp b} \,
\d^3 \bfk_{D} \,
\delta (\bfk_{D} \cdot \hat{\bfp}_c) \,
\delta (\hat s +\hat t +\hat u - 2m_c^2) \> \nonumber \\
& \times {\mathcal J}(z,\bfk_D) \, \Biggl\{ \sum_q
\left[ \left ( -\frac{k_{\perp\,a }}{M_p}  \right ) f_{1T}^{\perp\, q}(x_a, k_{\perp a})\cos\phi_a \,f_{\bar q/p}(x_b,
k_{\perp b}) \>
H^{\rm Inc}_{q\bar q\to Q\overline Q} (\hat s, \hat t , \hat u)\,
 D_{D/Q}(z,\bfk_D) \right]  \nonumber \\
& + \left[  \left ( -\frac{k_{\perp\,a }}{M_p}  \right ) f_{1T}^{\perp\, g \,(f)}(x_a, k_{\perp a})  \cos\phi_a\,  f_{g/p}(x_b,  k_{\perp b}) \>
H^{{\rm Inc}\,(f)}_{gg\to Q \overline Q}(\hat s, \hat t, \hat u)\,
 D_{D/Q}(z,\bfk_D)  \right .       \nonumber \\
& + \left .  \left ( -\frac{k_{\perp\,a }}{M_p}  \right ) f_{1T}^{\perp\, g \,(d)}(x_a, k_{\perp a})  \cos\phi_a\,  f_{g/p}(x_b,  k_{\perp b}) \>
H^{{\rm Inc}\,(d)}_{gg\to Q \overline Q}(\hat s, \hat t, \hat u)\,
 D_{D/Q}(z,\bfk_D) \right] \Biggr\}\,,
\label{eq:num-D}
\end{align}
where
\begin{align}
H^{{\rm Inc} \,(f)}_{gg\to c\bar c} & = H^{{\rm Inc}\,(f)}_{gg\to \bar c c} = -\frac{N_c}{4(N_c^2-1)}\,  \frac{1}{\tilde t \tilde u} \, \left (  \frac{\tilde t^2}{\tilde s^2}+ \frac{1}{N_c^2}\right ) \left ( \tilde t^2 + \tilde u^2 + 4 m_c^2 \tilde s - \frac{4 m_c^4 \tilde s^2}{\tilde t \tilde u }    \right )\, , \nonumber \\
H^{{\rm Inc} \,(d)}_{gg\to c\bar c} & =- H^{{\rm Inc}\,(d)}_{gg\to \bar c c} = -\frac{N_c}{4(N_c^2-1)}\,  \frac{1}{\tilde t \tilde u} \, \left (  \frac{\tilde t^2 - 2 \tilde u^2}{\tilde s^2}+ \frac{1}{N_c^2}\right ) \left ( \tilde t^2 + \tilde u^2 + 4 m_c^2 \tilde s - \frac{4 m_c^4 \tilde s^2}{\tilde t \tilde u }    \right )~.
\label{eq:Hfd}
\end{align}
These last two equations are in agreement with the hard partonic cross sections in Ref.~\cite{Kang:2008ih}, which have been  calculated in the twist-three approach.

\subsection{Numerical results}
\label{Dmes-res}
In addition to gluon TMD-PDFs, which, as already seen, contribute to $A_N$ for $p^\uparrow p \to J/\psi\,X$, in inclusive $D$ production one needs to consider quark TMD-PDFs and FFs as well. For the $\bm k_\perp$ dependence of the unpolarized quark distributions we assume the same simple Gaussian parametrization as in Eq.~(\ref{eq:unp-TMD}), with $\langle k_\perp^2 \rangle = 0.25$ GeV$^2$. For the unpolarized fragmentation functions we adopt a similar model, in which the dependences on $z$ and $k_{\perp D}$ are factorized,
\begin{equation}
 D_{D/Q}(z, k_{\perp D}) = D_{D/Q}(z) \,g(k_{\perp D})\,,
\end{equation}
with  $D_{D/Q}(z)$ being the collinear fragmentation function, for which we use the LO parametrization in Ref.~\cite{Kniehl:2006mw}, and $g(k_{\perp D})$ is a Gaussian function as in Eq.~(\ref{eq:unp-TMD}) with $\langle k_{\perp D}^2 \rangle = 0.2$  GeV$^2$~\cite{Anselmino:2004nk}\footnote{Notice that this value has been obtained for the light-quark FFs into a pion; we have checked that using larger values, up to 1 GeV$^2$, has a very tiny effect on SSA estimates.}, normalized in such a way that
\begin{equation}
\int \d^2 \bm k_{\perp D} \, D_{D/Q}(z, k_{\perp D}) = D_{D/Q}(z)~.
\end{equation}
We assume to have only one nonzero fragmentation function for $D$ mesons,
\begin{equation}
D_{D^0/c} (z)= D_{\bar D^0/\bar c}(z) =   D_{D^+/c}(z) = D_{D^-/\bar c} (z)\,,
\end{equation}
and all the other contributions are set to zero.

\begin{figure*}[t]
\begin{center}
{\includegraphics[trim =  50 60 190 550,clip,width = 8.cm]{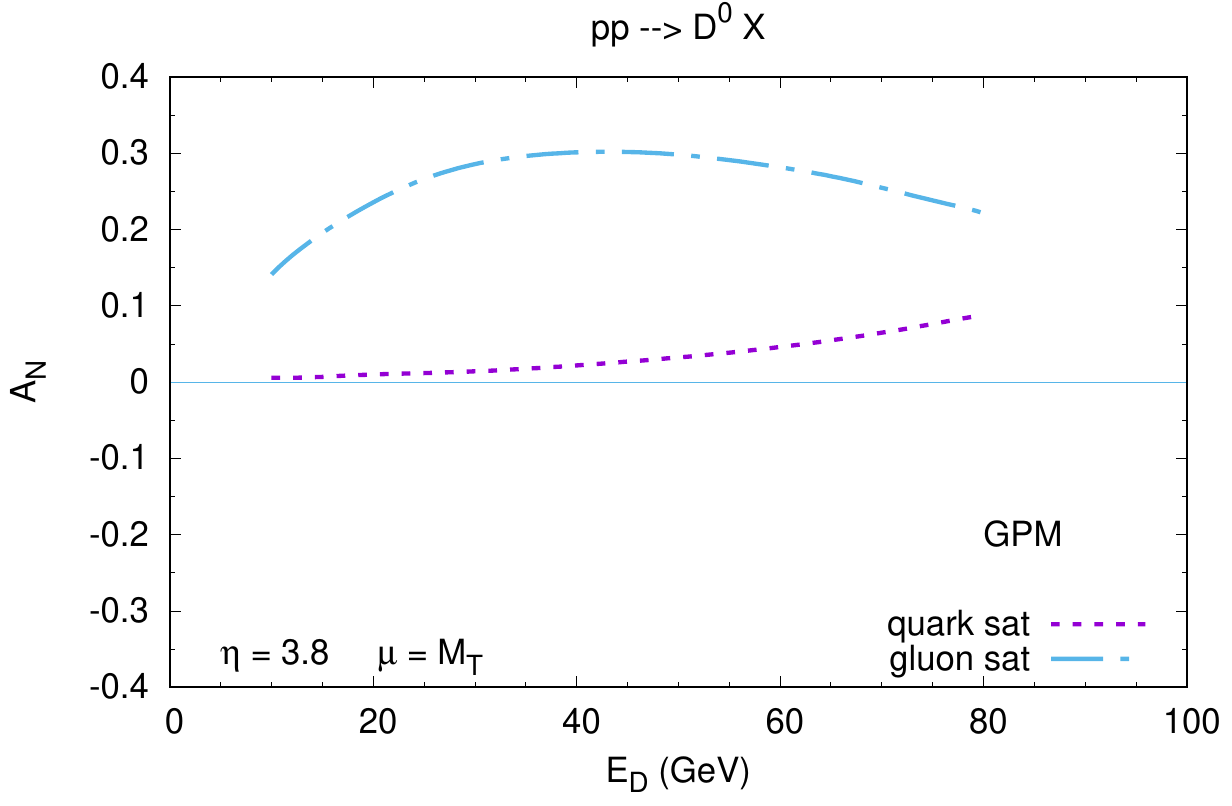}}
\hspace*{0.5cm}
{\includegraphics[trim =  50 60 190 550,clip,width = 8.cm]{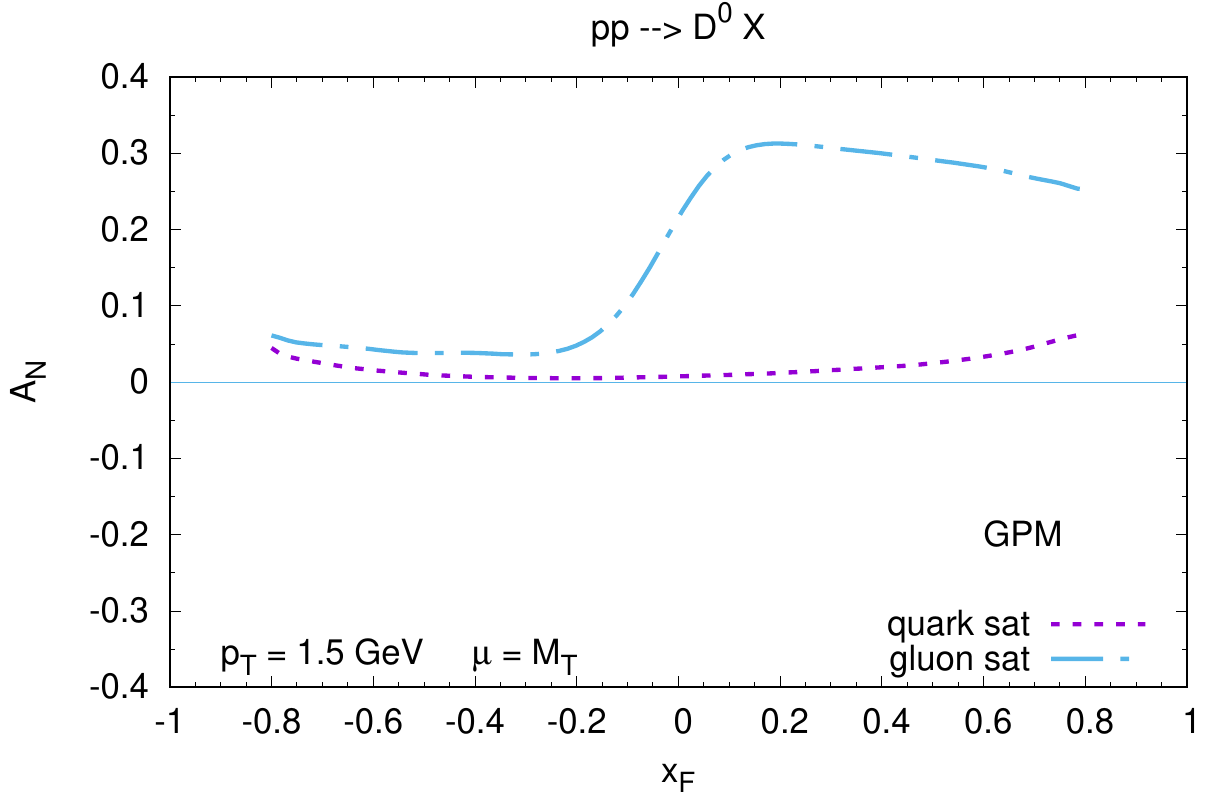}}
\end{center}
\caption{Quark and gluon contributions to the upper bounds (${\cal N}_{q,g}(x)=+1$) on $A_N$  for the process $p^\uparrow p\to D^0X$ calculated in the GPM approach at $\sqrt{s} = 200$ GeV: at fixed pseudorapidity, $\eta=3.8$, as a function of $E_D$  (left panel), and at fixed transverse momentum, $p_T = 1.5$ GeV, as a function of $x_F$ (right panel).  }
\label{fig:AN-GPM-D}
\end{figure*}

In the calculation of the upper bounds for the SSAs, we adopt for all quark and gluon Sivers functions the functional form  in Eq.~(\ref{eq:siv-par}) with ${\cal N}_{q,g}(x)=+1$ and $\rho =2/3$. Moreover, we take $\langle k_\perp^2\rangle = 1 $ GeV$^2$ for gluons  and $\langle k_\perp^2\rangle = 0.25 $ GeV$^2$ for quarks. The factorization scale is chosen to be equal to the transverse mass of the $D$ meson, $\mu= M_T$, with $M_D=1.869$ GeV (for the charm mass entering the hard scattering parts we adopt $m_c=1.3$~GeV). We note that the estimates for the denominators of the asymmetries are the same in both the GPM and the CGI-GPM approaches.

Our GPM results for $A_N$ in $p^\uparrow p\to D^0\,X$, which are the same as the ones for $p^\uparrow p\to \bar{D}^0\,X$, are presented in Fig.~\ref{fig:AN-GPM-D} in two different kinematic regions: at fixed pseudorapidity, $\eta=3.8$, as a function of $E_D$~\cite{Anselmino:2004nk} (left panel), and at fixed transverse momentum, $p_T = 1.5$ GeV, as a function of $x_F$ (right panel).

\begin{figure*}[t]
\begin{center}
\subfloat[] {\includegraphics[trim =  50 60 190 550,clip,width = 8.5cm]{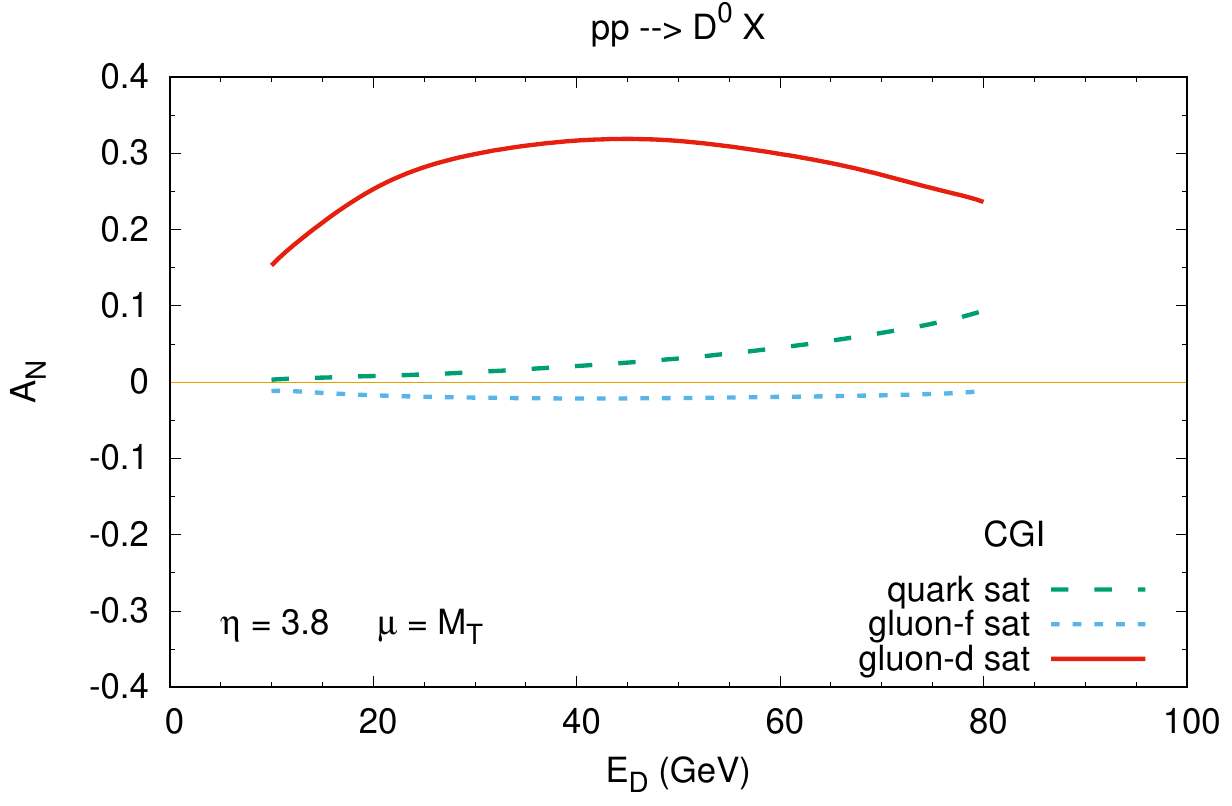}}
\hspace*{0.5cm}
\subfloat[] {\includegraphics[trim =  50 60 190 550,clip,width = 8.5cm]{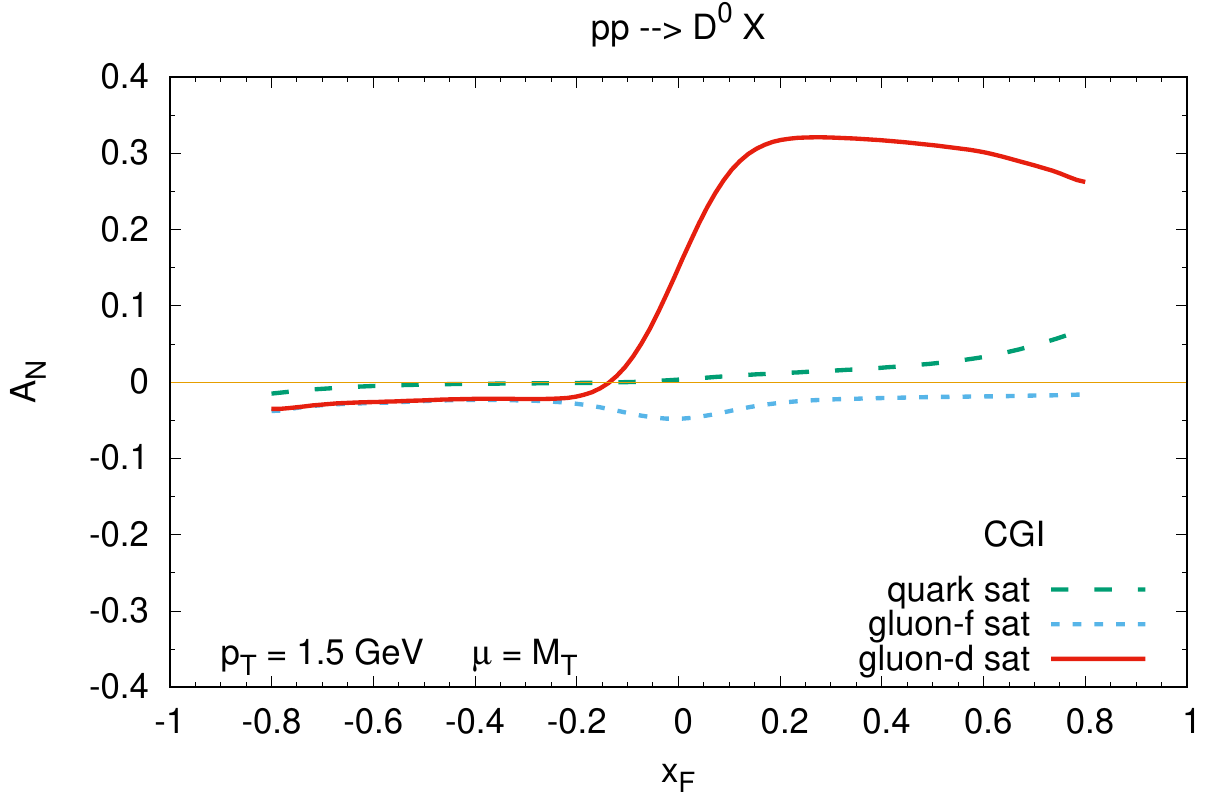}}

\subfloat[]  {\includegraphics[trim =  50 60 190 550,clip,width = 8.5cm]{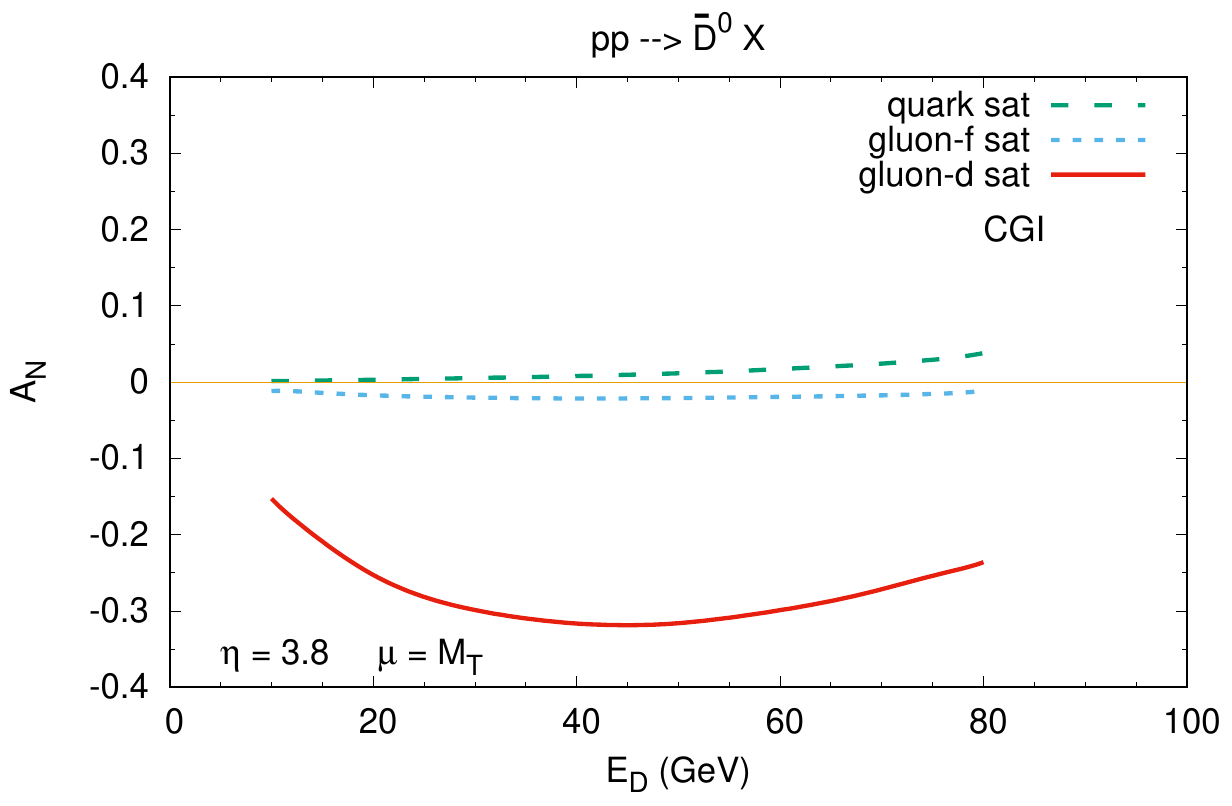}}\hspace*{0.5cm}
\subfloat[] {\includegraphics[trim =  50 60 190 550,clip,width = 8.5cm]{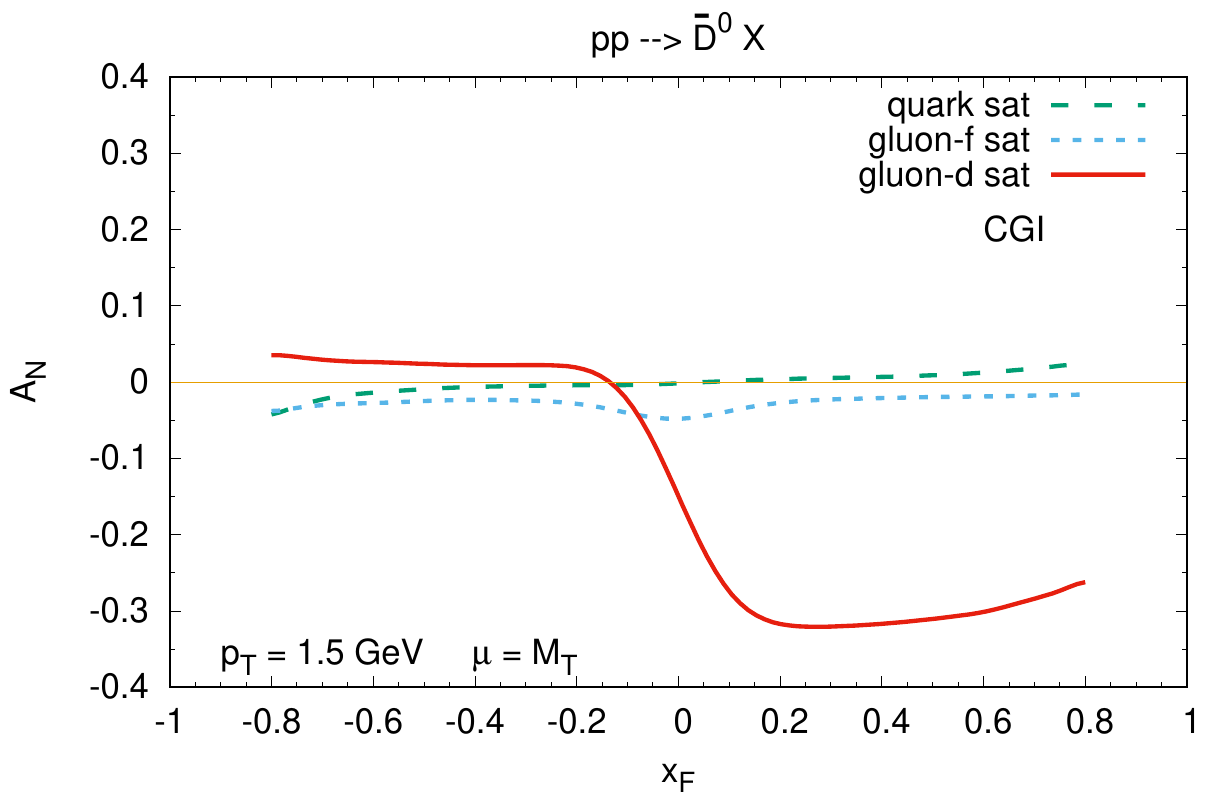}}
\end{center}
\caption{Upper bounds in size (that is taking ${\cal N}_{q,g}(x)=+1$) for the quark, $f$-type and $d$-type gluon Sivers contributions to $A_N$ in the process $p^\uparrow p\to D^0X$ calculated in the CGI-GPM approach at $\sqrt{s} = 200$ GeV: at fixed pseudorapidity,
$\eta=3.8$, as a function of $E_D$ (a), and at fixed transverse momentum, $p_T = 1.5$ GeV, as a function of $x_F$ (b). Analogous results for the process $p^\uparrow p\to \bar{D}^0\,X$ are presented in panels (c) and (d).}
\label{fig:AN-CGI-D}
\end{figure*}

One of the main results of the above calculation (analogous to what happens in the twist-three formalism) is that in the CGI-GPM approach, $A_N$ for $D^0$ meson is different from  $A_N$ for $\bar{D}^0$, as shown in Fig.~\ref{fig:AN-CGI-D}, where the same kinematic regions as in Fig.~\ref{fig:AN-GPM-D} have been considered.
We find that the quark contributions to $A_N$ in both models are almost negligible for values of the $D$ meson energy $E_D \le  40$ GeV and for $x_F \le 0.6$.
It is worth pointing out that adopting any of the GPM quark Sivers functions as extracted from data on azimuthal asymmetries in SIDIS processes would give an almost negligible contribution to $A_N$, leaving at work only the gluon Sivers effect.

The gluon contribution in the GPM is relatively large in size for $x_F \ge 0$ and in the whole considered range of $E_D$. For $D^0$ production, in the CGI-GPM, the $f$-type gluon Sivers effect is always quite small, while the $d$-type is similar to the $f$-type for $x_F<0$, and to the GPM for $x_F>0$. This can be understood by looking at Eqs.~(\ref{eq:Hfd}), where the $d$ and $f$-type hard functions differ only by one term. Indeed for negative $x_F$, $ \vert \tilde{u}\vert < \vert \tilde{t}\vert $ and the two hard functions give almost the same contribution. For positive $x_F $, $\vert \tilde{u}\vert $ becomes large and relevant and the $d$-type contribution in Eq.~(\ref{eq:Hfd}) becomes positive. Notice that the small size of all the asymmetries in the negative $x_F$ region is due to the integration over the Sivers azimuthal phase.
We also point out that the different behavior of the $f$ and $d$-type hard functions under the $c \leftrightarrow \bar c$ charge conjugation is not relevant since the FF for a $\bar{c}$ into a $D^0$ is taken to be zero.

By comparison of the upper and lower panels of Fig.~\ref{fig:AN-CGI-D}, it is clear that there is no difference between the $f$-type gluon asymmetries for $D^0$ and $\bar{D}^0$ production, while we find a tiny difference for the quark Sivers asymmetries and a change of sign for the $d$-type contributions, see Eq.~(\ref{eq:Hfd}). These findings imply that for positive $x_F$, a sizable difference in the asymmetries for $D^0$ and $\bar{D}^0$ would validate the CGI-GPM framework (or, equivalently, disprove the GPM), and, at the same time, would provide an indication of the size of the unknown Sivers functions $f_{1T}^{\perp\,g\,(d)}$. On the contrary, if $f_{1T}^{\perp\,g\,(d)}$ is very small, the GPM and CGI-GPM would predict the same asymmetry for $D^0$ and $\bar{D}^0$, making it impossible to distinguish between the two models. Furthermore, if we consider the SSA for the production of both $D^0$ and $\bar D^0$, the following relation holds
\begin{equation}
 A_N(D^0+\bar D^0) = \frac{1}{2} \left [ A_N(D^0) + A_N(\bar{D}^0) \right ]\,,
\end{equation}
which is valid in both models because the unpolarized cross sections for $D^0$  and $\bar{D}^0$ are the same. In the GPM this asymmetry would be the same as for $D^0$ or $\bar D^0$ production, while in the CGI-GPM it would receive a (small) contribution only from $f_{1T}^{\perp\,g\,(f)}$, since the one from $f_{1T}^{\perp\,g\,(d)}$ cancels in the sum. In other words a sizable $A_N(D^0+\bar D^0)$ at forward rapidities could be expected only within the GPM approach.

Finally, we note that the simultaneous study of $A_N$ for inclusive $D$ and $\bar D$ meson production has been already suggested in order to disentangle the two trigluon correlation functions in the twist-three formalism~\cite{Kang:2008ih,Koike:2011mb}. Notice that our estimates cannot be compared directly with those presented in Ref.~\cite{Koike:2011mb}, since here we have only considered a maximized scenario, without any attempt to constrain the gluon Sivers parameterizations. What we can only point out is that, in both the GPM and the CGI-GPM approaches, the asymmetry in the backward region cannot be sizeable, due to the integration over the azimuthal phases. This is in contrast to what happens in the twist-three formalism, where one could get $A_N$ values of the order of 30\% for $x_F<0$.

Concerning the comparison with the experimental results, namely the RHIC data~\cite{Liu:2009zzw,Nagashima:2017mpm,Aidala:2017pum} from the PHENIX Collaboration, one has to recall that, in order to do it, $A_N$ for $D$ mesons has to be converted into $A_N$ for $\mu$ production, taking into account the $D\to\mu $ kinematics. This would be a very important analysis with different potential outcomes: $i)$ discriminating among different approaches (TMD vs.~twist-three scheme); $ii)$ discriminating among TMD models (GPM vs.~CGI-GPM); $iii)$ putting, within a TMD scheme, some constraints on the gluon Sivers functions.

\section{Conclusions}
\label{concl}
In this paper we have performed a detailed analysis, within a TMD factorization scheme, of SSAs for inclusive hadronic processes characterized by one large energy scale and dominated by gluon-gluon fusion contributions, with two important aims: from one side we have addressed the role of the TMD gluon Sivers function, still largely unknown and from the other one we have studied its process dependence, intimately connected to the universality issue.
To this end, we have considered two inclusive processes, namely $J/\psi$ and $D$ meson production in $pp$ collisions, for which gluon initiated subprocesses are expected to be dominating, extending to the gluon sector the inclusion of initial- and final-state color interactions, responsible for the process dependence of the TMDs. We have then presented theoretical estimates obtained by adopting both a generalized parton model approach with inclusion of spin and transverse momentum effects and its color-gauge invariant extension, still based on a partonic interpretation, which includes also ISI and FSI effects via a one-gluon exchange approximation.

Concerning charmonium production, adopting the color-singlet model we have shown that with the inclusion of TMD effects, and taking into account the uncertainty coming from the choice of the factorization scale, the theoretical estimates are able to reproduce the central rapidity RHIC data reasonably well, at least for $p_T$ values lower than 2~GeV. On the other hand the available SSA data are still not precise enough to discriminate among the two models or to give any robust constraint on the GSF.

Moving to $D$ meson production, in the CGI-GPM approach we have shown the emergence of two independent gluon Sivers functions, according to how the color is neutralized. As a clear signature these two GSFs enter differently in $D^0$ and  $\bar D^0$ mesons, providing a tool to disentangle them and at the same time to check the validity of the GPM approach (where we have only one GSF) or its CGI version.

To this end, present available SSA data (requiring a suitable conversion of $A_N$ for $D$ meson to $\mu$ meson production) and future experimental results could be extremely important to check the validity of the approaches, put some constraints on the gluon Sivers function and test its universality properties.

It is worth mentioning that $J/\psi$ and $D$ meson production could also be studied in $\pi p^\uparrow$ collisions at COMPASS. While this would be less sensitive (as compared to RHIC) to gluons, it could nevertheless provide useful insights into the $J/\psi$-production mechanism through quarks and antiquarks in the TMD approach.

The richness of present and forthcoming experimental activities together with their complementary and educated phenomenological analyses are opening a new era in learning about the inner mechanisms behind transverse SSAs, and as a by-product, about some challenging features of QCD. Within a TMD scheme, these studies also provide a powerful tool to get information on gluon TMDs, and in particular on the role of color exchange and its impact on the process dependence of the gluon Sivers function.

\acknowledgments
We would like to thank Leonard Gamberg and Jean-Philippe Lansberg for useful discussions; Bernd Kniehl, Gustav Kramer and Ingo Schienbein for providing us with the code for the $D$ meson fragmentation functions; Christine Aidala, Jeongsu Bok and Stephen Pate for useful information on the experimental data. C.P.~is supported by the European Research Council (ERC) under the European Union's Horizon 2020 research and innovation programme (grant agreement No.~647981, 3DSPIN).

\appendix
\section{Unpolarized cross section for $g\, g \to J/\psi \, g $}
\label{app:unp-J}
\begin{figure}[t]
\begin{center}
\includegraphics[trim= 55 650 40 80,clip,width=16cm]{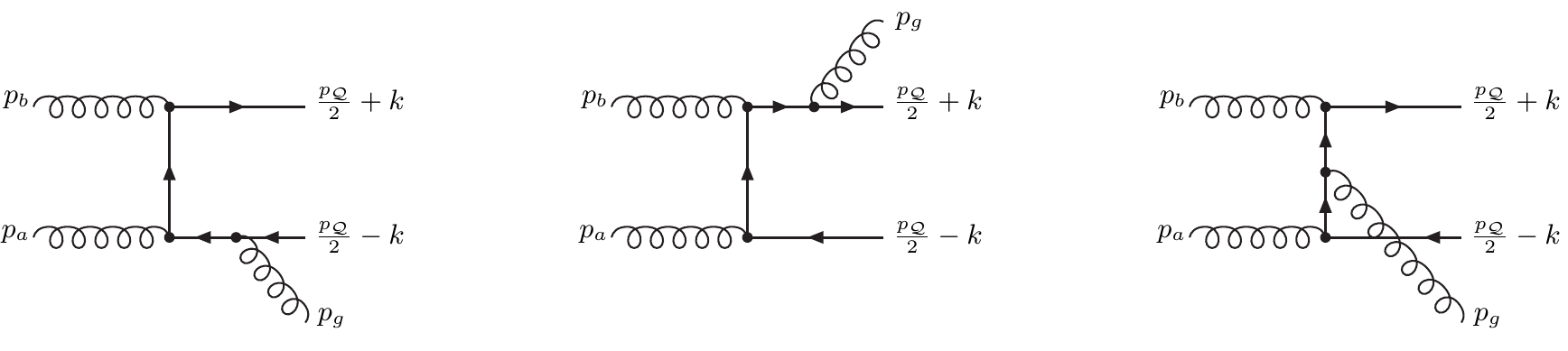}
\end{center}
\caption{Feynman diagrams for the process $gg\to Q \overline Q  [^3S^{(1)}_1] \, g$ in the CSM at
LO in perturbative QCD. The corresponding crossed diagrams, included in the calculation, are not shown. They can be obtained by reversing the fermion lines and replacing $k\leftrightarrow -k$.}
\label{fig:gglo}
\end{figure}

The scattering amplitude ${\cal M}$ for the partonic process  $g(p_a)\,{+}\,g(p_b)\,\to\, Q \overline Q  [^3S^{(1)}_1] (p_{\cal Q})\,{+}\,g(p_g)$ at LO in the CSM can be written in the form \cite{Baier:1983va}:
\begin{eqnarray}
{\cal{M}} (p_a, p_b; p_{\cal Q}, p_g) &  = &   \frac{1}{4 \sqrt{\pi\, M }}\, R_0(0)\, {\rm Tr}\left [O(p_a, p_b; p_{\cal Q}, 0)\, (\pqs - M) \,\epss_{\lambda_{\cal Q}}^*(p_{\cal Q}) \right ]\,,
\label{eq:3S1a}
\end{eqnarray}
where $R_0(0)$ is the radial wave function of the bound state evaluated at the origin and $\varepsilon_{\lambda_{\cal Q} }(p_{\cal Q})$ is its polarization vector. The operator  $O(p_a,p_b; p_{\cal Q}, k)$ is calculated from the Feynman diagrams in Fig.~\ref{fig:gglo}, where $k$ is half the relative momentum of the two outgoing quarks forming the bound state, which we take to be zero in our nonrelativistic approximation. We adopt the notation $O(0) \equiv O(p_a,p_b; p_{\cal Q}, 0)$. Moreover,  we  write
\begin{equation}
O(p_a, p_b; p_{\cal Q}, k) = \sum_{m=1}^6 {\cal C}_m \,O_m(p_a, p_b; p_{\cal Q}, k)\,,
\label{eq:O1}
\end{equation}
where we have separated the color factors ${\cal C}_m$ from the rest of the Feynman amplitudes $O_m$. Explicitly, from the three diagrams in Fig.~\ref{fig:gglo}, one gets:
\begin{align}
O_1 & =  4 g_s^3 \,\varepsilon_{\lambda_a}^{\mu}(p_a)\, \varepsilon_{\lambda_b}^{\nu}(p_b)\,\varepsilon^{\rho \,*}_{\lambda_g}(p_g)\,  \gamma_\nu\,  \frac{\pqs + 2 \ks - 2 {\ps_b} +
2 M_Q}{(p_{\cal Q}-2p_b+2k)^2-4 M_Q^2}\,\gamma_\mu \, \frac{-\pqs + 2 \ks - 2 {\pgs} + 2 M_Q}{(p_{\cal Q}+2 p_g-2k)^2-4 M_Q^2} \,\gamma_\rho\,, \nonumber\\
O_2 & =   4 g_s^3 \,\varepsilon_{\lambda_a}^{\mu}(p_a)\, \varepsilon_{\lambda_b}^{\nu}(p_b)\, \varepsilon^{\rho\,*}_{\lambda_g}(p_g) \, \gamma_\rho\, \frac{\pqs + 2 \pgs + 2 {\ks} +
2 M_Q}{(p_{\cal Q} + 2 p_g + 2k)^2-4 M_Q^2} \,\gamma_\nu \,\frac{-\pqs + 2 \ps_a + 2 {\ks} + 2 M_Q}{(p_{\cal Q}-2p_a-2k)^2-4 M_Q^2}\,\gamma_\mu \,,\nonumber \\
O_3 & =   4 g_s^3 \,\varepsilon_{\lambda_a}^{\mu}(p_a)\, \varepsilon_{\lambda_b}^{\nu}(p_b)\, \varepsilon^{\rho\,*}_{\lambda_g}(p_g) \,  \gamma_\nu\, \frac{\pqs - 2 \ps_b + 2 {\ks} +
2 M_Q}{(p_{\cal Q} - 2 p_b + 2k)^2-4 M_Q^2} \,\gamma_\rho \,\frac{-\pqs + 2 \ps_a + 2 {\ks} + 2 M_Q}{(p_{\cal Q}-2p_a-2k)^2-4 M_Q^2}\,\gamma_\mu \,,
\label{eq:O}
\end{align}
and
\begin{align}
{\cal C}_1 =\sum_{i,j}\,\langle 3i;\bar{3}j\vert 1\rangle \, (t^b t^at^c)_{ij} \,, \qquad {\cal C}_2 =\sum_{i,j}\,\langle 3i;\bar{3}j\vert 1\rangle \, (t^ct^b t^a)_{ij} \,, \qquad {\cal C}_3 =\sum_{i,j}\,\langle 3i;\bar{3}j\vert 1\rangle \, (t^bt^c t^a)_{ij} \,,
\label{eq:Ci0}
\end{align}
where the sum is taken over the colors of the outgoing quark and antiquark and $t^a$ are the $SU(3)$ generators in the fundamental representation, normalized according to Tr$(t^at^b) = \delta^{ab}/2$. The $SU(3)$ Clebsch-Gordan coefficients,
\begin{equation}
\langle 3i;\bar{3}j\vert 1\rangle = \frac{\delta^{ij}}{\sqrt{N_c}}\,,
\label{eq:cs}
\end{equation}
with $N_c$ being the number of colors, project out the color-singlet configuration. By substituting Eq.~(\ref{eq:cs})  in Eq.~(\ref{eq:Ci0}), we obtain
\begin{equation}
{\cal C}_1 = {\cal C}_2 = \frac{1}{4 \sqrt{N_c}}\, (d_{abc} - i f_{abc})\,,
\qquad  {\cal C}_3 = \frac{1}{4 \sqrt{N_c}}\, (d_{abc} + i f_{abc})~.
\end{equation}
 The other color factors ${\cal C}_{4,5,6}$ can be obtained from  ${\cal C}_{1,2,3}$, respectively,  by exchanging
 $a \leftrightarrow b$. Therefore we find
\begin{equation}
{\cal C}_4 = {\cal C}_5 = \frac{1}{4 \sqrt{N_c}}\, (d_{abc} + i f_{abc})\,,\qquad {\cal C}_6 = \frac{1}{4 \sqrt{N_c}}\, (d_{abc} - i f_{abc})~.
\end{equation}
The operators $O_m$, with $m \ge 4$,  can be obtained from the ones in Eq.~(\ref{eq:O}) by applying crossing relations. Since the relations
\begin{eqnarray}
{\rm Tr}\left [O_1(0)\, (\pqs - M) \,\epss_{\lambda_{\cal Q}}^*(p_{\cal Q}) \right ] & = &{\rm Tr}\left [O_4(0)\, (\pqs - M) \,\epss_{\lambda_{\cal Q}}^*(p_{\cal Q}) \right ]\,, \nonumber \\
{\rm Tr}\left [O_2(0)\, (\pqs - M) \,\epss_{\lambda_{\cal Q}}^*(p_{\cal Q}) \right ] & = &{\rm Tr}\left [O_5(0)\, (\pqs - M) \,\epss_{\lambda_{\cal Q}}^*(p_{\cal Q}) \right ]\,, \nonumber \\
{\rm Tr}\left [O_3(0)\, (\pqs - M) \,\epss_{\lambda_{\cal Q}}^*(p_{\cal Q}) \right ] & = &{\rm Tr}\left [O_6(0)\, (\pqs - M) \,\epss_{\lambda_{\cal Q}}^*(p_{\cal Q}) \right ]\,,
\end{eqnarray}
hold, in the sum of all the contributions to the amplitude, by adding first diagrams 1 and 4, 2 and 5, and 3 and 6, we single out  the same, symmetric, combinations of color factors, namely
\begin{equation}
{\cal C}_1 + {\cal C}_4 = {\cal C}_2 + {\cal C}_5 = {\cal C}_3 + {\cal C}_6 = \frac{1}{2\sqrt{N_c}}\, d_{abc}~.
\label{eq:CFJ}
\end{equation}
Hence Eq.~(\ref{eq:3S1a}) can be written as
\begin{eqnarray}
{\cal{M}}  (p_a, p_b; p_{\cal Q}, p_g) & = & \frac{1}{4 \sqrt{\pi\, M }}\, R_0(0)\,  \frac{1}{2\sqrt{N_c}}\, d_{abc}\, {\rm Tr}\left [ \sum_{m=1}^3 O_{m}(0)\, (\pqs - M) \,\epss_{\lambda_{\cal Q}}^*(p_{\cal Q}) \right ]\,,
\label{eq:3S1}
\end{eqnarray}
with
\begin{align}
 \hspace{-0.3cm}\sum_{m=1}^3 O_m(0) & =  {g_s^3}\,\varepsilon_{\lambda_a}^{\mu}(p_a)\, \varepsilon_{\lambda_b}^{\nu}(p_b)\, \varepsilon^{\rho\, *}_{\lambda_g}(p_g)\, \left [ \frac{\gamma_\nu
 (\pqs -2\ps_b + M) \,\gamma_\mu\,(-\pqs  - 2 {\pgs} + M)\,\gamma_\rho}{(\hat s -M^2)(\hat u-M^2)} \right . \nonumber  \\
&  \left . \, + \, \frac{\gamma_\rho  (\pqs +2\pgs + M) \,\gamma_\nu\, (-\pqs  + 2 {\ps}_a + M)\,\gamma_\mu}{(\hat s -M^2)(\hat t-M^2)} +
\frac{\gamma_\nu (\pqs -2\ps_b + M) \,\gamma_\rho\,  (-\pqs  + 2 {\ps}_a + M)\,\gamma_\mu}{(\hat t -M^2)(\hat u-M^2)} \right ]\,,
\label{eq:O0}
\end{align}
where  have introduced the Mandelstam variables
\begin{equation}
\hat s = (p_a+p_b)^2\equiv (p_{\cal Q} + p_g)^2\,, \qquad \hat t = (p_a- p_{\cal Q})^2\,,\qquad \hat u = (p_b- p_{\cal Q})^2~.
\end{equation}
The unpolarized partonic cross section for the process $g g\to J/\psi \,g$ is given by
\begin{equation}
\frac{\d\hat \sigma}{\d\hat t} = \frac{1}{16\pi\hat s^2 }\, \overline{|{\cal M}|^2}\,,
\end{equation}
where an average is understood over the initial gluon polarizations and colors, and a sum over the final ones. When summing over the polarizations of the on-shell gluons, care must be taken to consider only the physical (transverse) polarization states. This  can be achieved through the following relations: 
\begin{align}
\sum_{\lambda_{a}}\varepsilon_{\lambda_{a}}^{\mu}(p_{a})\, \varepsilon_{\lambda_{a}}^{\mu^\prime *}(p_{a}) & =
- \left [ g^{\mu\mu^\prime} - \frac{2}{\hat s}\, \left  (p_a^\mu \, p_b^{\mu^\prime} + p_a^{\mu^\prime} p_b^{\mu}\right  ) \right ] \, ,\nonumber \\
 \sum_{\lambda_{b}}\varepsilon_{\lambda_{b}}^{\nu}(p_{b}) \,\varepsilon_{\lambda_{b}}^{\nu^\prime *}(p_{b}) & =  - \left [ g^{\nu\nu^\prime} - \frac{2}{\hat s}\, \left  (p_a^\nu \, p_b^{\nu^\prime} + p_a^{\nu^\prime} p_b^{\nu}\right  ) \right ]  \,, \nonumber \\
 \sum_{\lambda_{g}}\varepsilon_{\lambda_{g}}^{\rho}(p_{g}) \,\varepsilon_{\lambda_{g}}^{\rho^\prime *}(p_{g}) & =  - \left [ g^{\rho\rho^\prime} - \frac{2}{\hat u}\, \left  (p_a^\rho \, p_g^{\rho^\prime} + p_a^{\rho^\prime} p_g^{\rho}\right  ) \right ] \,,
\end{align}
while the sum over the polarization states of the $J/\psi$ is performed by using the identity
\begin{equation}
\sum_{\lambda_{\cal Q}}\varepsilon_{\lambda_{\cal Q}}^{\alpha}(p_{\cal Q}) \,\varepsilon_{\lambda_{\cal Q}}^{\beta \,*}(p_{\cal Q}) =
-g^{\alpha\beta} + \frac{p_{\cal Q}^\alpha p_{\cal Q}^\beta}{M^2}~.
\end{equation}
The final result is given in Eqs.~(\ref{eq:unp-cs}) and (\ref{eq:HU}), where we have taken $N_c=3$.

\section{Color factors in $p^\uparrow p \to D\, X$ within the CGI-GPM framework}
\label{app:CGI-D}

\begin{figure}[t]
\begin{center}
\includegraphics[trim= 80 700 50 50,clip,width=17cm]{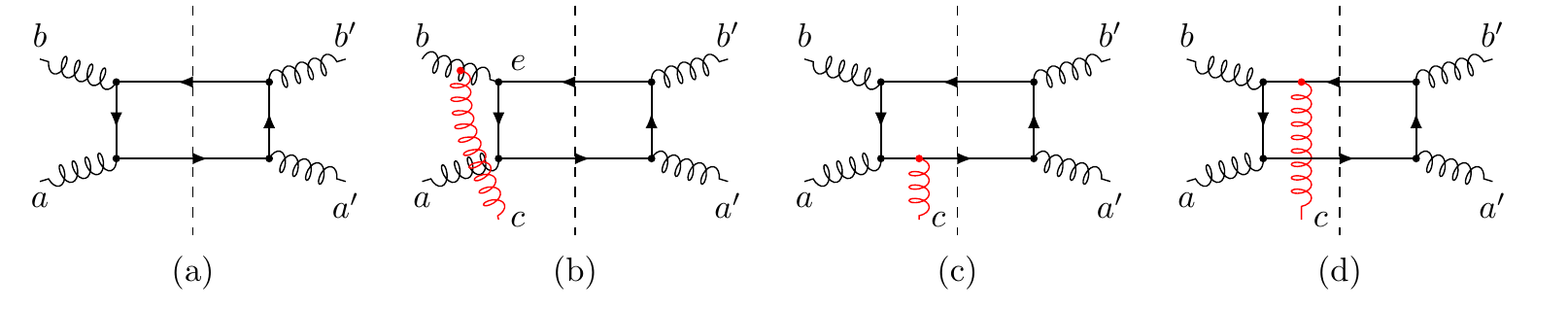}
\end{center}
\caption{Diagrams contributing to the color factors $C_{U}$ (a), $C_{I}^{\left(f/d \right)}$ (b),
$C_{F_{c}}^{\left( f/d \right)}$ (c) and $C_{F_{d}}^{\left(f /d \right)}$ (d) for the process $gg\to c \bar c$.}
\label{fig:ccbar example}
\end{figure}

In this appendix we provide an explicit calculation of the color factors needed for the estimation of the SSAs for
$p^\uparrow p \to D \,X$ in the CGI-GPM approach. We limit our discussion to the first diagram in Table~\ref{tab:gg2ccb}, the extension to the remaining ones being straightforward. We begin with the color factor for the unpolarized amplitude squared in Fig.~\ref{fig:ccbar example}(a), which is given by
\begin{equation}
\begin{aligned}C_{U} & =\frac{\delta_{bb'}\delta_{aa'}}{\left(N_{c}^{2}-1\right)^{2}}\mathrm{Tr}\left[t^{a}t^{a'}t^{b'}t^{b}\right]=\frac{1}{4N_{c}}.\end{aligned}
\end{equation}
The diagram in Fig.~\ref{fig:ccbar example}(b)  accounts for the ISI, described by an additional eikonal gluon attached to the initial-state gluon that comes from the unpolarized proton. From the CGI-GPM rules in Fig.~\ref{fig:Feynman}, $C_{I}^{\left(f\right)}$ and $C_{I}^{\left(d\right)}$ are calculated as follows:
\begin{equation}
\begin{aligned}C_{I}^{\left(f\right)} & =\frac{\delta_{bb'}}{N_{c}^{2}-1}\mathcal{T}_{aa'}^{c}T_{ce}^{b}\mathrm{Tr}\left[t^{e}t^{b'}t^{a'}t^{a}\right],\\
 & =-\frac{N_{c}}{8\left(N_{c}^{2}-1\right)},
\end{aligned}
\end{equation}
and
\begin{equation}
\begin{aligned}C_{I}^{\left(d\right)} & =\frac{\delta_{bb'}}{N_{c}^{2}-1}\mathcal{D}_{aa'}^{c}T_{ce}^{b}\mathrm{Tr}\left[t^{e}t^{b'}t^{a'}t^{a}\right],\\
 & =\frac{N_{c}}{8\left(N_{c}^{2}-1\right)}\,,
\end{aligned}
\end{equation}
where ${{\cal T}}_{aa'}^{c}$  and ${\cal D}^c_{aa^\prime}$ are respectively the $f$ and $d$-type projectors defined
in Eqs.~(\ref{eq:cp}) and (\ref{eq:norm-cp}), while $T^b_{ce}\equiv -i f_{bce}$. Similarly, the color factors $C_{F_{c}}^{\left(f\right)}$ and $C_{F_{c}}^{\left(d\right)}$ related to the FSI of the outgoing charm quark, are obtained by adding an eikonal gluon to the lower quark line as depicted in Fig.~\ref{fig:ccbar example}(c), from which we find
\begin{equation}
\begin{aligned}C_{F_{c}}^{\left(f\right)} & =\frac{\delta_{bb'}}{N_{c}^{2}-1}\mathcal{T}_{aa'}^{c}\mathrm{Tr}\left[t^{b}t^{b'}t^{a'}t^{c}t^{a}\right]=\frac{1}{8N_{c}},\end{aligned}
\end{equation}
and
\begin{equation}
\begin{aligned}C_{F_{c}}^{\left(d\right)} & =\frac{\delta_{bb'}}{N_{c}^{2}-1}\mathcal{D}_{aa'}^{c}\mathrm{Tr}\left[t^{b}t^{b'}t^{a'}t^{c}t^{a}\right]=\frac{1}{8N_{c}}.\end{aligned}
\end{equation}
Finally, one computes the color factors $C_{F_{d}}^{\left(f\right)}$ and $C_{F_{d}}^{\left(d\right)}$ from Fig.~\ref{fig:ccbar example}(d), where the eikonal gluon is now attached to the upper antiquark line. We get
\begin{equation}
\begin{aligned}C_{F_{d}}^{\left(f\right)} & =-\frac{\delta_{bb'}}{N_{c}^{2}-1}\mathcal{T}_{aa'}^{c}\mathrm{Tr}\left[t^{b}t^{c}t^{b'}t^{a'}t^{a}\right]\\
 & =-\frac{1}{8N_{c}\left(N_{c}^{2}-1\right)},
\end{aligned}
\end{equation}
where the minus sign in the first line stems from the antiquark propagator, see the color rule in Fig.~\ref{fig:Feynman}(c). Likewise,
\begin{equation}
\begin{aligned}C_{F_{d}}^{\left(d\right)} & =-\frac{\delta_{bb'}}{N_{c}^{2}-1}\mathcal{D}_{aa'}^{c}\mathrm{Tr}\left[t^{b}t^{c}t^{b'}t^{a'}t^{a}\right]\\
 & =\frac{1}{8N_{c}\left(N_{c}^{2}-1\right)}~.
\end{aligned}
\end{equation}

\end{document}